\documentclass[aps,prd,reprint,superscriptaddress]{revtex4-1}

\usepackage{graphicx}				
\usepackage{color}
\definecolor{orange}{rgb}{1,0.5,0}
\definecolor{gray}{rgb}{0.4,0.4,0.4}
\definecolor{midnight}{rgb}{0.1,0.1,0.6}
\usepackage[normalem]{ulem}
\newcommand{\added}[1]{{\color{red}#1}}
\newcommand{\removed}[1]{{\color{midnight}\sout{#1}}}

\renewcommand{\added}[1]{#1}
\renewcommand{\removed}[1]{}

\newcommand{\Ka}{K$\alpha$\ }
\newcommand{\Kb}{K$\beta$\ }
\newcommand{\La}{L$\alpha$\ }

\newcommand{\Ll}{L\emph{l}}
\newcommand{\diff}{\mathrm{d}}

\newcommand{\affilNISTPML}{\affiliation{U.~S. National Institute of Standards and Technology, Quantum Electromagnetics Division, 325 Broadway MS 687, Boulder, Colorado, USA 80305}}
\newcommand{\affilNISTITL}{\affiliation{U.~S. National Institute of Standards and Technology, Applied \& Computational Mathematics Division, 325 Broadway MS 771, Boulder, Colorado, USA 80305}}
\newcommand{\affilNISTRadPhys}{\affiliation{U.~S. National Institute of Standards and Technology, Radiation Physics Division, 100 Bureau Drive, Stop 8460, Gaithersburg, Maryland USA 20899}}

\newcommand{\affilCU}{\affiliation{Department of Physics, University of Colorado, Boulder, Colorado, USA 80309}}
\newcommand{\affilTheiss}{\affiliation{\added{Theiss Research, 7411 Eads Ave, La Jolla, California USA 92037}}}

\begin{document}
\title{A Reassessment of Absolute Energies of the X-ray L Lines of Lanthanide Metals}
\author{J. W. Fowler}
\email[Correspondence should be addressed to: ]{joe.fowler@nist.gov} \affilNISTPML

\author{B.~K. Alpert}\affilNISTITL
\author{D.~A. Bennett}\affilNISTPML
\author{W.~B. Doriese}\affilNISTPML
\author{J.~D. Gard}\affilNISTPML\affilCU
\author{G.~C. Hilton}\affilNISTPML
\author{L.~T. Hudson}\affilNISTRadPhys
\author{Y.-I. Joe}\affilNISTPML
\author{K.~M. Morgan}\affilNISTPML
\author{G. C. O'Neil}\affilNISTPML
\author{C.~D. Reintsema}\affilNISTPML
\author{D.~R. Schmidt}\affilNISTPML
\author{D.~S. Swetz}\affilNISTPML
\author{C.~I. Szabo}\affilNISTRadPhys\affilTheiss
\author{J.~N. Ullom}\affilNISTPML\affilCU
\date{\today}							

\begin{abstract}
We introduce a new technique for determining x-ray fluorescence line energies and widths, and we present measurements made with this technique of 22 x-ray L lines from lanthanide-series elements. The technique uses arrays of transition-edge sensors, microcalorimeters with high energy-resolving power that simultaneously observe both calibrated x-ray standards and the x-ray emission lines under study. The uncertainty in absolute line energies is generally less than 0.4\,eV in the energy range of 4.5\,keV to 7.5\,keV\@. Of the seventeen line energies of neodymium, samarium, and holmium, thirteen are found to be consistent with the available x-ray reference data measured after 1990; only two of the four lines for which reference data predate 1980, however, are consistent with our results. Five lines of terbium are measured with uncertainties that improve on those of existing data by factors of two or more.
These results eliminate a significant discrepancy between measured and calculated x-ray line energies for the terbium \Ll\ line (5.551\,keV).
The line widths are also measured, with uncertainties of 0.6\,eV or less on the full-width at half-maximum in most cases.
These measurements were made with an array of approximately one hundred superconducting x-ray microcalorimeters, each sensitive to an energy band from 1\,keV to 8\,keV. No energy-dispersive spectrometer has previously been used for absolute-energy estimation at this level of accuracy.  Future spectrometers, with superior linearity and energy resolution, will allow us to improve on these results and expand the measurements to more elements and a wider range of line energies.
\end{abstract}

\pacs{}

\maketitle

\section{Introduction}

Since the Nobel-Prize winning discovery of ``characteristic R\"ontgen radiation'' by Barkla \citep{Stephenson:1967}, the emission and absorption of x rays by each of the elements has been a topic of interest.  The body of information that encompasses properties such as emitted x-ray energies, lineshapes, fluorescence yields, absorption probabilities, and absorption edges of the elements is now known as \emph{x-ray fundamental parameters}.  These parameters are of practical significance because they facilitate compositional analysis of complex materials.  They are also a potent test of atomic theory, because x-ray emission and absorption are caused by electronic transitions between quantized, element-specific orbitals. For example, Barkla's characteristic radiation is caused by the filling of core holes by electrons from less tightly bound orbitals with the concomitant emission of an x ray.  The energy of the emitted x rays corresponds to the difference in the binding energies of the final and initial electronic states.
In general this emission is accompanied by less intense satellite lines due to other mutiple-electron excitations that depend on the details of the excitation and decay processes.

Traditionally, absolute x-ray energies have been measured via diffraction techniques, in which a crystalline material is illuminated with a tightly collimated beam of  x rays from the sample to be measured.  If the measurement geometry is well understood and if the lattice spacing of the crystal is known, then the angles at which intense diffraction features appear reveal the x-ray wavelength.
While a small fraction of these results have part-per-million  (ppm) accuracy, most have far larger uncertainties, largely because high-resolution, wavelength-dispersive measurements require a difficult and exacting technique; very few labs are attempting to make updated measurements. 

Tabulations of x-ray lines were compiled as recently as 2003 \citep{Deslattes:2003} and 2007 \citep{Zschornack:2007wu}. The \citet{Deslattes:2003} review, also known as NIST Standard Reference Database 128, is the official United States publication on x-ray line energies.  It is the first full collection of x-ray K and L transition energies to combine measurements and theoretical calculations.  Both recent reviews rely heavily on measurements performed decades earlier, particularly on those compiled by \citet{Bearden:1967tg}. Many of the tabulated values were measured before x-ray wavelength standards were traceable to the definition of the meter~\cite{Deslattes:1973}. If some older measurements have systematic errors due to the conversion from early ``x units'' or any other cause, the effects would be nearly impossible to reconstruct many decades later. The authors published the best available value for each transition, but there remain numerous opportunities for modern measurements to offer meaningful improvements in such tabulations.  With the passage of time, new needs for x-ray fundamental parameters have emerged~\citep{Beckhoff:2012um}, along with opportunities to measure fundamental parameters in new ways.

Here we introduce a technique to determine x-ray line energies and widths.  This technique is based on simultaneous measurement of multiple lines previously determined to great accuracy through the crystal diffraction method, as well as other lines whose spectral features are of interest but are less well known.  The measurements are performed with energy-dispersive detectors that have sufficient spectral resolution to discern most spectral features and smoothly varying but \emph{a priori} unknown calibration curves.   The well-known spectral features are used to estimate the calibration curve of the detector, which then establishes the energies of the unknown spectral features.

This type of measurement has attractions for determination of x-ray fundamental parameters.  Energy-dispersive detectors respond to photons over a broad range of energies so that many x-ray lines can be measured simultaneously.  Energy-dispersive detectors are also much more efficient than diffractive instruments and thus are well suited to measurement of the faintest spectral features.  Nonetheless, there are challenges.  Detector response can be---and in this case is---a nonlinear function of energy.  Further, the availability of suitable calibration features is constrained both by nature and by the extent of prior work.  Hence, this manuscript shows both the feasibility of and also the current limits on metrology with a sparse, nonlinear ruler.

Measurement with a non-ideal ruler is a common problem in x-ray science.  Spectral features in astrophysical plasmas are shifted and broadened by high velocities and temperatures.  Absolute energies and widths are therefore useful as kinematic probes.  Recent observations of the Perseus galaxy cluster by the Hitomi satellite enabled the quantification of a velocity gradient across the cluster core~\citep{Aharonian:2016jq}.  The uncertainty on the velocity gradient is largely determined by a 1\,eV absolute-energy uncertainty for spectral features near 6.5\,keV.  The \emph{Athena} x-ray observatory, a large project now in preparation, has a more ambitious absolute-energy specification of 0.4\,eV~\citep{Ravera:2014hv}.  X-ray spectral features produced by exotic atoms---atoms in which an electron is replaced by a muon, pion, kaon, or other exotic particle---are also shifted and broadened compared to their purely electronic counterparts.  The details of these spectral features depend on both quantum electrodynamics (QED) and the strong nuclear force, and their measurement continues to provide valuable tests of these theories~\citep{Gotta:2015,Anagnostopoulos:2003,Gal:2013}.  X-ray emission from highly charged ions (HCIs) similarly offers tests of QED~\added{\cite{Amaro:2012, Amaro:2013kv, Kubicek:2014gz}.
HCI emission measurements of helium-like ions have been made recently with energy-dispersive x-ray microcalorimeters at energies near 30\,keV~\cite{Beiersdorfer:2009}, below 14\,keV~\cite{Thorn:2009}, and near 1\,keV~\cite{Gillaspy:2011eo}.}
The absolute-energy uncertainties described in this manuscript are highly representative of the uncertainties that could be achieved in measurements of x rays from astrophysical, exotic-atom, or HCI sources by the same technique.

In this work, the energy-dispersive spectrometer consisted of 160 superconducting transition-edge sensors (TESs).  The sensors and the multiplexing electronics based on superconducting quantum interference device (SQUID) amplifiers are described elsewhere \citep{Ullom:2015kp}.  The core requirement of the individual sensors is an energy resolution of a few eV for x rays in the range of 4\,keV to 8\,keV, without which it would be difficult to identify, locate, or measure the shape of x-ray emission lines.  The TES's response to deposited energy is broadly understood, though it cannot now be computed at the quantitative level required for precision metrology~\citep{Ullom:2015kp}.  The use of many separate sensors in a pixelated array provides equally many independent measurements of the same spectral features, a crucial element in the assessment of error limits. We attempt here the first absolute-energy calibration of TES microcalorimeters with carefully studied systematic uncertainties and with achieved uncertainties better than 0.4\,eV over a broad energy range of 4.5\,keV to 7.5\,keV\@.

The most accurately measured x-ray features in this range are the K-shell transitions of the 3d transition metals.  These transitions span 4\,keV to 9\,keV; both line energies and detailed line shapes have previously been determined (e.g., by \citet{Holzer:1997ts}).  The most intense features in these complex spectra are located with 2\,ppm accuracy.  In contrast, the L-shell transitions of the lanthanides are much less well known; in several cases, existing uncertainties are at the level of 200\,ppm.   The L-shell lanthanide transitions span the same energy range as those of the 3d K-shells, so these two sets of elements provide a candidate combination of standard ruler and unknown.  

Existing measurements of L-shell lanthanide x-ray fundamental parameters date largely to the 1970s~\citep{Bearden:1967tg,Gokhale:1969wy,Gokhale:1970ve,Nigam:1973vp,Shrivastava:1977jd}.  Some, such as the terbium L lines, were measured much earlier~\citep{Coster:1922fw,Sakellaridis:1953tw} and transferred into later tables.
More recently, the use of the lanthanides and other rare earths in electronics, optoelectronics, glasses, lighting, and permanent magnets has expanded greatly.  Consequently, there is increased interest in analyzing lanthanide-bearing materials and ores by a variety of techniques including x-ray fluorescence analysis (XRF)~\citep{DAngelo:2001dm, Orescanin:2006ej, Taam:2013uu, Kirsanov:2015fi}.  Quantitative composition analysis by XRF requires knowledge of parameters such as line energies and shapes~\citep{Goldstein:1992}.  This knowledge is particularly critical when one uses common semiconductor detector systems whose spectral resolution is insufficient to fully separate relevant x-ray spectral features, for example x-ray lines from different lanthanides of interest and lines from the geologically more common 3d transition metals~\citep{Kirsanov:2015fi}.  Uncertainty in line centers results in either systematic or statistical uncertainty when one fits overlapping spectra to extract peak areas, depending on whether the center energies are treated as fixed or free parameters, respectively \citep{Hoover:2013}.   The relatively poor state of knowledge of lanthanide fundamental parameters and the growing technological significance of these materials motivates our work.


Here, we present both our new technique and our initial measurements of 22 L-series x-ray lines of four lanthanide metals. We describe the spectrometer (Section~\ref{sec:spectrometer}), the measurement technique and data analysis (Section~\ref{sec:measurements}), and the calibration procedure (Section~\ref{sec:calibration}). In Section~\ref{sec:results}, we present the results on line energies from metals where prior reference measurements were both relatively good (most lines of neodymium, samarium, holmium) and poor (terbium). In the first category, we find agreement with reference data for all of the lines. In the second category, we offer improved estimates of the line energies with uncertainties several times smaller than those in the best, earlier data.
We also present measurements of the widths of the emission lines, for which reference data are incomplete or inconsistent, and in some cases consist of theory alone (\citet{Zschornack:2007wu} and references therein).

\section{The Transition-Edge-Sensor Spectrometer} \label{sec:spectrometer}

The spectrometer consists of an array of 160 transition-edge sensor (TES) microcalorimeter detectors, operated at a temperature of approximately 100\,mK\@. The sensors effectively act as separate, independent spectrometers. They are illuminated with an x-ray flux that produces up to 100 photons per second in each. The electrical signal initiated by the absorption of a photon is digitized, recorded, and analyzed to produce an energy estimate with resolving power of at least 1000. Unlike a wavelength-dispersive spectrometer, each sensor in a microcalorimeter array detects photons throughout the full energy band.

The TES array is cooled without liquid cryogens by a commercially produced cryostat that employs both a mechanical pulse-tube cryocooler to hold a large volume at 3\,K and an adiabatic demagnetization refrigerator to reach 100\,mK in the microcalorimeters and 65\,mK in their supporting electronics. The array is situated behind a vacuum window and a series of aluminum-on-polymer filters that block visible and infrared radiation. The x-ray transmission through the window and filters is approximately 65\,\% at 6\,keV, with most of the loss in a 10\,$\mu$m-thick aluminum filter on the 65\,mK stage. The absorption of x rays in the sensors is approximately 75\,\% at 6\,keV. Because the filters' transmission rises with increasing energy across the band of interest (4.5\,keV to 7.5\,keV), but the detectors' absorption falls, the resulting quantum efficiency is not a strong function of energy; it is between 40\,\% and 50\,\% throughout the band. The largest variation in quantum efficiency over any 100\,eV interval in that range is a factor of 1.02 at the 4.5\,keV end. Item \ref{systematic:QE} in Section~\ref{sec:systematics} describes the small systematic effect of variations in quantum efficiency.

Each x-ray sensor consists of a thermalizing absorber and a thermometer.  A layer of bismuth 2.5\,$\mu$m thick serves as the absorber.  A molybdenum-copper superconducting bilayer 300\,nm thick sits underneath the bismuth and acts as the thermometer.  The absorber and thermometer are deposited on a micromachined membrane for thermal isolation; each is approximately 350\,$\mu$m square.  The SiN$_\mathrm{x}$ membrane is silicon-enhanced relative to stoichiometric Si$_3$N$_4$ to reduce stresses. The thermometer is voltage-biased into the superconducting-to-normal transition where its resistance is a strong function of temperature.  When the energy of an x-ray photon is deposited into the absorber, the TES temperature and resistance rise, creating a rapid decrease in the current through the TES. The thermal energy is quickly conducted away, and the TES returns to its quiescent state in a matter of milliseconds~\citep{Irwin-Hilton:2005}. The resulting negative current pulse is read out by a SQUID amplifier.

To minimize the wiring complexity and the heat load onto the coldest stages of the spectrometer, we employ a time-division multiplexed (TDM) readout system, composed of very low-noise SQUID amplifiers~\citep{Doriese:2015il}.  Eight amplifier channels read out the full array of 160 sensors. Each single channel samples the current through twenty separate sensors in succession, repeating one full cycle through all twenty every 12.8\,$\mu$s. Inductive, low-pass, anti-aliasing filters on the TES circuits reduce their response above the Nyquist frequency of 39\,kHz. 
A SQUID is an intrinsically nonlinear current amplifier, so the TDM channels employ flux-locked loops with active feedback to linearize the current sensing. The feedback current serves as the linearized estimate of the TES current.

\begin{figure}
\includegraphics[width=\linewidth]{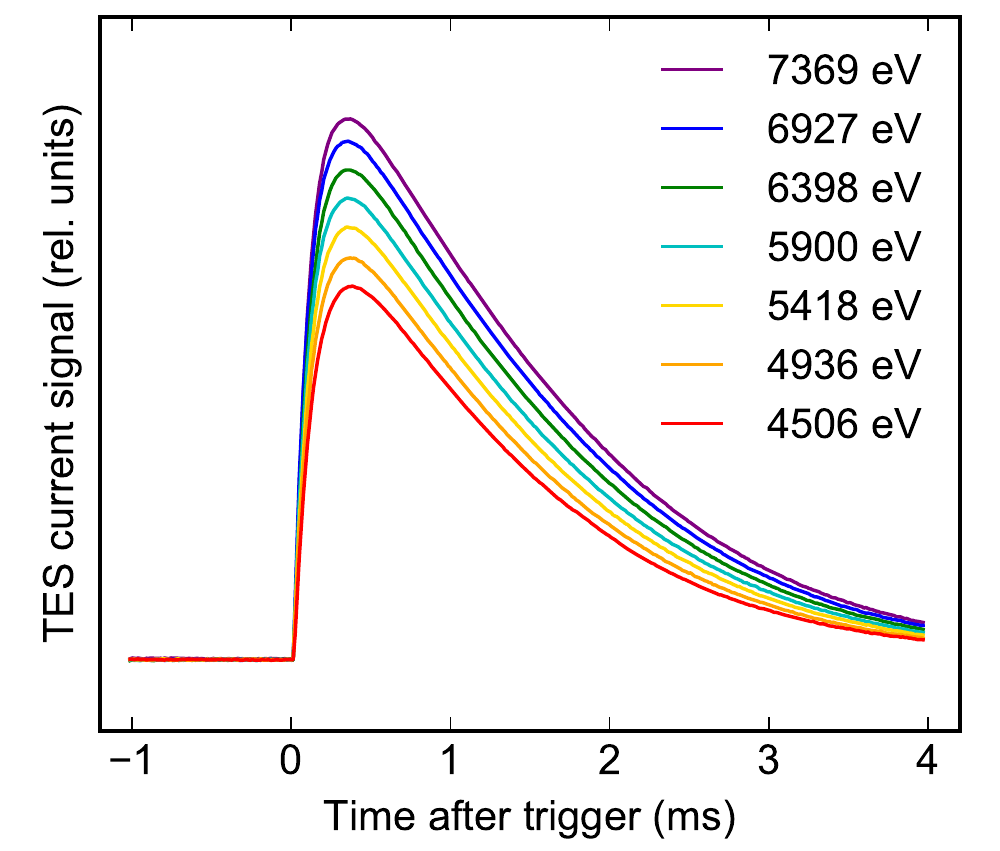}
  \caption{ \label{fig:raw_traces}
 Seven individual x-ray pulse records from the same TES.  The records correspond to the x-ray fluorescence lines Tb L$\beta_2$, Co K$\alpha$, Fe K$\alpha$, Mn K$\alpha$,  Cr K$\alpha$, Ti K$\beta$, and Ti \Ka (top to bottom). The seven records include detector noise, though it is too small to see at this scale.}
\end{figure}

\begin{figure}[t]
\includegraphics[width=\linewidth]{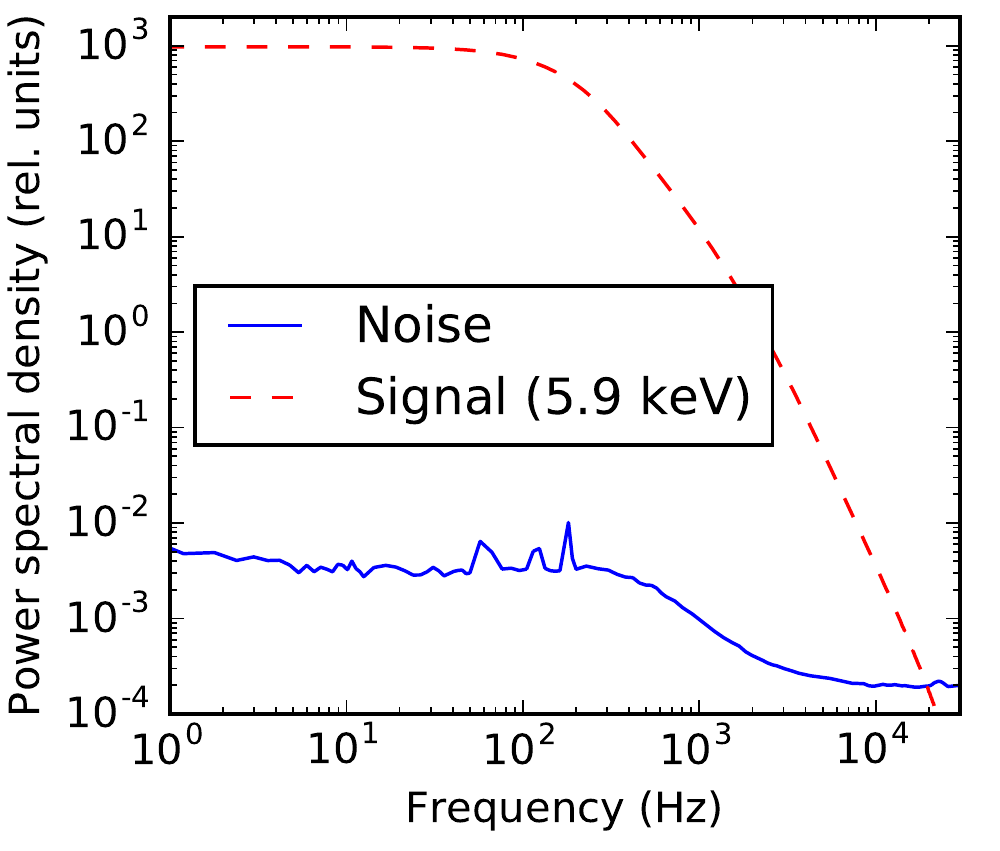}
\caption{ \label{fig:power_spectra}
The power spectral density of the typical noise and of the signal from a 5900 eV x ray. Both spectra are averaged across a representative sample of TES detectors. The signal-to-noise ratio is very high throughout the band below 1 kHz.}
\end{figure}

The spectrometer produces 160 data streams, each consisting of samples of the current in one sensor at the rate of 78 kilosamples per second per sensor. A triggering algorithm in the data-acquisition computer examines the data streams to identify photon events, generating records to be stored to disk whenever the current slope exceeds a predefined threshold. The very low noise of the sensors allows us to set this threshold at a level of approximately 100\,eV to 300\,eV, which is well below the energies of interest here. For this measurement, each data record consists of 390 samples. These samples include 1\,ms of the pre-pulse (``baseline'') period and the first 4\, ms of the photon pulse.
Figure~\ref{fig:raw_traces} shows representative pulse records for x rays of seven different energies, recorded by the same TES\@. Increases on the (uncalibrated) vertical axis correspond to transient \emph{decreases} in the bias current flowing through the TES. The shapes of TES pulses are similar at all energies, with only small departures from exact proportionality. 
Figure~\ref{fig:power_spectra} shows the power-spectral density of both noise and a typical single-pulse signal, indicating the extremely high signal-to-noise ratio in the band of interest, below approximately 1\,kHz.

We have operated similar spectrometers at x-ray beamlines of several synchrotrons, including the National Synchrotron Light Source at Brookhaven National Laboratory \citep{Ullom:2014er}, the Advanced Photon Source at Argonne National Laboratory, and the Stanford Synchrotron Radiation Lightsource at the Stanford Linear Accelerator Center. A thorough review of the physics, the readout, and the wide range of applications of superconducting microcalorimeter spectrometers is  given by~\citet{Ullom:2015kp}.

\section{X-ray fluorescence line measurements} \label{sec:measurements}

\subsection{X-ray fluorescence source and targets}

The commercial x-ray source used to produce fluorescence accelerates electrons from a cathode through 20\,kV towards a rhodium primary target. Brehmsstrahlung and characteristic fluorescence x rays produced in the rhodium illuminated the secondary target, which contained one or more metal foils, the elements of interest. For this measurement, we tuned the cathode current to achieve a flux of approximately 15 x-ray photons per second onto each TES, which were located at a distance of 13\,cm from the secondary target and had an active area of 0.1\,mm$^2$ apiece. The volume containing the secondary targets and the window of the spectrometer was held under weak vacuum, with an absolute pressure of roughly 8\,kPa.

The secondary targets were pieces of 99.9\,\% pure metals purchased from chemical-supply companies. For calibration, we used metallic titanium, chromium, manganese, iron, and cobalt ($Z=22,24,25,26,27$). The \Ka and K$\beta$ emission-line energies and shapes of these transition metals are well established \cite{Holzer:1997ts, Chantler:2006va, Chantler:2013wp}. The elements of primary interest in this study are four from the lanthanide series: neodymium, samarium, terbium, and holmium ($Z=60,62,65,67$). Three secondary targets, each 1\,cm square, were assembled from pieces of five or more metals. Table~\ref{tab:targets} gives the composition of the three targets, which was governed by the need to measure many emission lines in one observation while avoiding overlap between any lines of interest.

\begin{table}
\begin{tabular}{rl}
Target & Elements \\ \hline
1 & Ti, Cr, Mn, Fe, Co \\
2 & Ti, Cr, Mn, Fe, Co, Sm, Ho \\
3 & Ti, Cr, Mn, Fe, Co, Nd, Tb, Ho \\
\end{tabular}
\caption{ \label{tab:targets}
Contents of the three multi-metal targets.}
\end{table}

We performed initial, exploratory measurements to verify that the relative intensities of the main emission lines from each element in each target were similar and adjusted the size of the metal pieces in each target to ensure this.  The relative intensities of lines varied by a factor of approximately two across the spectrometer array, owing to geometrical factors in the design of the targets, the supporting structure, and the stainless steel parts that enclose the vacuum space around the spectrometer window. In the case of a few sensors, iron emission from Target~3 was too dim to use for calibration, so they were dropped from the analysis. With these exceptions, all elements could be observed in the spectra of all TESs.

\begin{figure*}
\includegraphics[width=\linewidth]{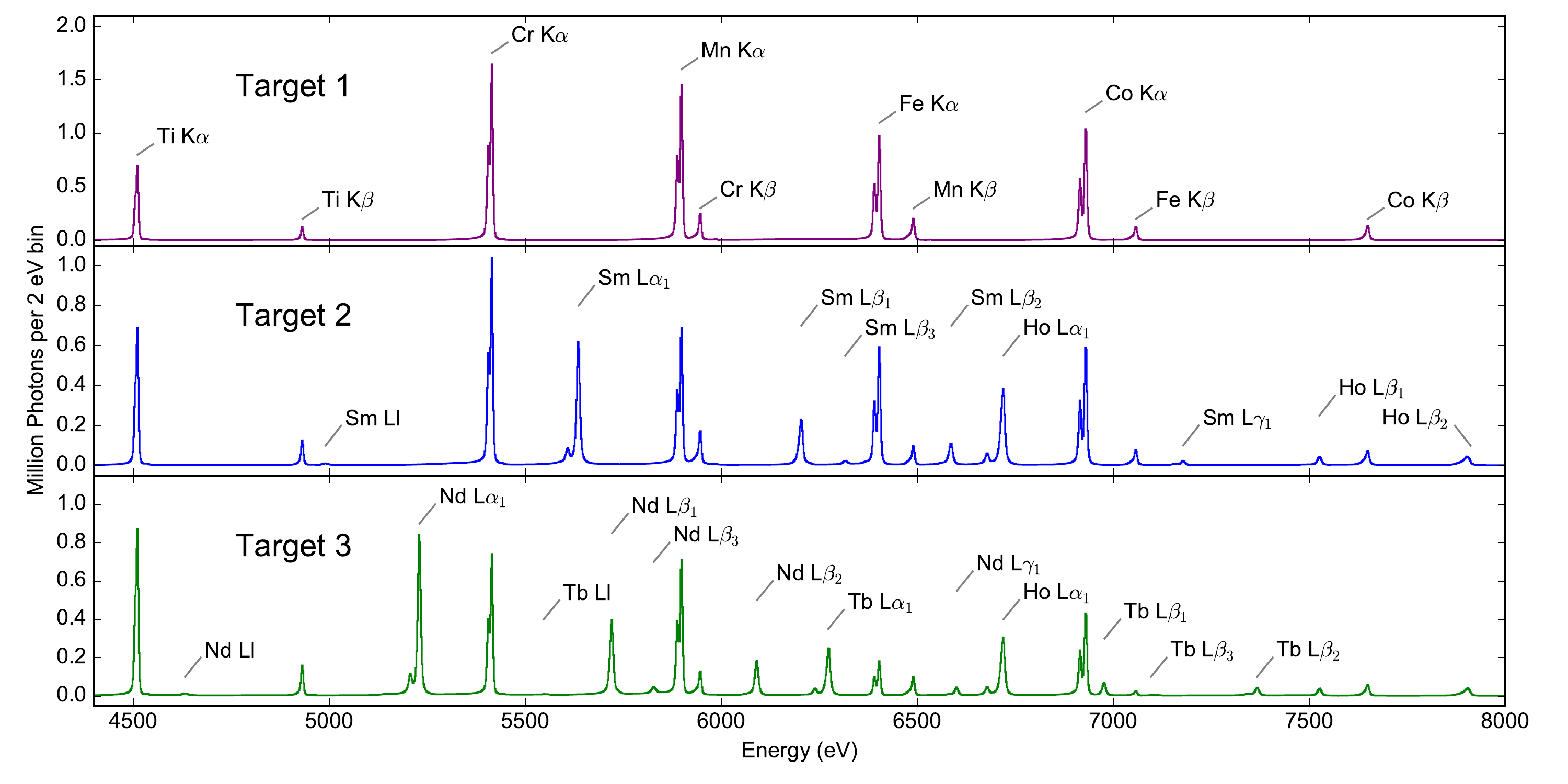}
\caption{ \label{fig:all_spectra}
The measured spectra of targets 1, 2, and 3 (top to bottom), in the primary spectral region of interest:  4.4 to 8.0 keV\@. The lines visible here are from the elements listed in Table~\ref{tab:targets}, including the K lines of five transition metals and the L lines of four elements from the lanthanide series. These spectra consist of data from all TESs that were successfully calibrated.
}
\end{figure*}

The data presented here consist of three observations: one measurement of 12\,h to 14\,h on each of  Targets 1 to 3. These generated a total of roughly 180\,GB  of raw data in the form of 250 million records of TES current pulses. After the analysis steps described in this and the next section, the observations yielded the wide-band x-ray energy spectra shown in Figure~\ref{fig:all_spectra}. While the vast majority of x rays---including those in all the emission lines under study---were in the 4\,keV to 8\,keV energy range, other very dim lines between 1.4\,keV and 4\,keV were also observed. Lines throughout the wider energy range can be used for absolute-energy calibration, as described in Section~\ref{sec:calibration}.

\subsection{Pulse selection and estimation of pulse heights} \label{sec:pulse_analysis}

The millions of pulse records are analyzed by techniques described more thoroughly by \citet{Fowler:2015is}, which we only summarize here. A small set of pulse-summary quantities are used to select only ``clean'' single-pulse events. The selection eliminates piled-up (multi-pulse) records, photons that arrive before the energy of the previous photon is fully dissipated, and any other transient problems. Approximately 80\,\% of records remain after these cuts.

Each clean record is reduced to a single pulse-height estimator by  linear optimal filtering. This means that a specific weighting of the 390 samples in a record is chosen to yield the minimum-variance, unbiased estimator for the pulse height. This weighting is statistically optimal under several assumptions that are known to be approximately but not exactly true: that pulses are transient departures from a strictly constant baseline level; that the noise is an additive, stationary, multivariate Gaussian process independent of the signal level; and that all pulses at any energy are proportional to a single standard pulse shape.

The filtered pulse height is subject to a number of small, unwanted systematic dependences on variables such as the baseline level and the photon's exact arrival time at the sub-sample level.  Slow variations in the  baseline level track changes in the temperature of the cryogenic thermal bath, which in turn produce changes in the sensor gain at the level of $10^{-3}$. We correct the gain by a factor linear in the observed baseline level, selecting a correction factor for each sensor that minimizes the information entropy of its resulting energy spectrum~\citep{Fowler:2015is}.  The pulse-height bias that depends on the exact photon arrival time relative to the 12.8\,$\mu$s sampling clock is also a function of the photon energy, unfortunately.  We reduce the bias by imposing a low-pass filter on the basic optimal filter (a one-pole filter with $f_{3\mathrm{dB}}=5$\,kHz) and then correct for the remaining bias by an algorithm that matches the most prominent peaks as a function of arrival time, and that linearly interpolates these corrections at intermediate energies.  The corrections for gain drift and subsample arrival-time are typically equivalent to changes of 5\,eV or less. After these adjustments, each record has been converted to a minimum-noise, bias-corrected estimate of its pulse height.

\section{Absolute Calibration of the Spectrometer} \label{sec:calibration}

The goal of an absolute-energy calibration is to determine one function, $f_i$, for each TES detector, such that $E=f_i(P)$ is the best possible estimate of the energy of a photon detected in sensor number $i$, when the filtered and corrected pulse height is $P$. The calibration procedure is critical to the success of metrological measurements; as such, we have explored its details more carefully than in previous work with microcalorimeters.

The TES-based spectrometer is not currently amenable to an energy calibration based on an understanding of the physics of the sensors. This physics is the subject of active research, and many recent developments have improved our understanding (for a summary, see \citet{Ullom:2015kp}). The resistance of a TES is a complicated function of its temperature, the local magnetic field, and the current passing through it.  As a result, the calibration curve cannot be estimated from first principles with an accuracy comparable to the part-per-thousand resolving power of the devices.  
To reach the full potential of a TES spectrometer, then, we must embark on a careful program of absolute calibration based directly on the pulse data. In this program, well established x-ray emission lines are interleaved with the lines being studied, so that the former can anchor a calibration function from pulse heights to photon energies for characterizing the latter. In this section we describe the anchor lines, the procedure for selecting a curve, and our estimates of uncertainty and sources of systematic variance.

\subsection{Calibration lines from transition metals}  \label{sec:cal_transition_metals}

The well known fluorescence lines that anchor our calibration are the K$\alpha$ and K$\beta$ lines of the elements in Target~1: titanium, chromium, manganese, iron, and cobalt. (Vanadium, with $Z=23$, is not used because its K lines are within 15\,eV of the Ti K$\beta$ and Cr \Ka lines. While the TES can easily resolve lines at this separation, the near-overlap of the lines would interfere with the accurate fitting of line energies.) Absolutely calibrated wavelength-dispersive spectrometers have measured the K-line energies and line shapes of these transition metals. The titanium \Ka and \Kb models come from \citet{Chantler:2006va} and \citet{Chantler:2013wp}, respectively.  The other four elements are from \citet{Holzer:1997ts}.  We assume these shapes and especially these line energies to be correct to within their stated uncertainties (typically 0.01\,eV for the \Ka lines and a few times larger for the K$\beta$ lines).

To generate each calibration anchor point in $P$-$E$ space, we perform a fit of our measured spectra to the appropriate line shapes in the model over a narrow energy range. This step uses a maximum-likelihood fit, which has much smaller biases \citep{Fowler:2014JLTP} than a simpler fit minimizing Neyman's or Pearson's $\chi^2$. The calibration point equates the known line energy with the pulse height at which the line peaks (or strictly speaking, where it would have peaked in a hypothetical measurement with perfect energy resolution).

The model to be fit must account for the intrinsic energy response of the TESs. A first approximation to the TES energy-response function is a Gaussian, which we characterize by its full width at half maximum. This Gaussian width is typically between 4\,eV and 6.5\,eV for these data. 
The instrumental broadening increases slowly across the energy range of interest: the relevant noise is roughly constant in pulse-height units, while the nonlinear calibration function converts this constant pulse-height error into a slowly growing energy error.
The specific design of these TESs, however, exhibits an additional feature caused by the bismuth absorbers: a one-sided exponential tail to low energies. The full energy-response function is thus taken to be a {Bortels function} \citep{Bortels:1987}, the convolution of a Gaussian with the normalized sum of a one-sided exponential and a delta function. This energy-response function has three parameters: the Gaussian energy resolution, $\delta E$; the scale length, $\lambda$, of the low-energy exponential tail; and the fraction, $f_{\mathrm{tail}}$, of the integral that is due to the exponential tail rather than the delta function. The model has five additional parameters: the offset and slope of a (locally) linear conversion from pulse height to energy; the line intensity; and the slope and intercept of a linear function representing background counts per spectral bin.

\begin{figure}
\begin{center}
\parbox[b]{\linewidth}{
	\includegraphics[width=\linewidth,keepaspectratio]{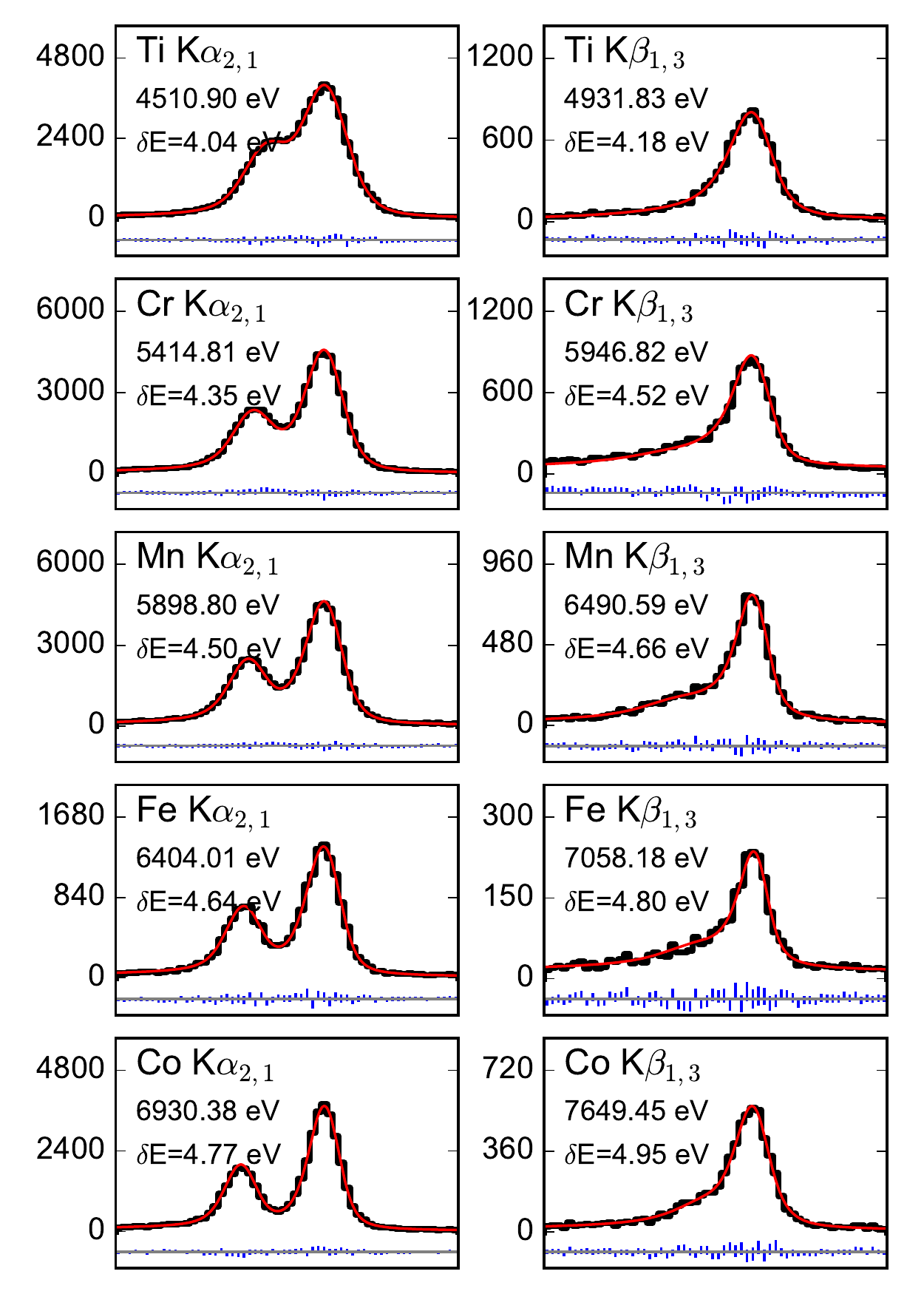}
}
\end{center}
\caption{Absolute calibration is performed on each detector separately, starting with maximum-likelihood fits of the measured spectra to the known fluorescence-line shapes of the K$\alpha_{1,2}$ and K$\beta_{1,3}$ lines of the five calibration elements Ti, Cr, Mn, Fe, and Co. Examples of these ten fits are shown for a single detector: the measured spectra as histograms (black) and the best fit as a solid overlaid curve (red). Each horizontal span covers a fractional range of 0.65\,\% in filtered pulse height, divided into 60 bins.
From each fit we extract the line's peak location in pulse-height units and---for \Ka fits only---the measurement's energy resolution (see Table~\ref{tab:resolutions}). Typical statistical uncertainties from the fits are roughly one part in $10^5$ on the peak locations and from 0.05\,eV to 0.10\,eV on the energy resolution (FWHM) for \Ka fits. The vertical bars beneath each spectrum indicate the extent of the 1$\sigma$ range around the residual (data minus fit).
} 
\label{fig:one_channel_calibration}
\end{figure}

The left panels of Figure~\ref{fig:one_channel_calibration} show examples of these eight-parameter fits to the \Ka doublets of the five calibration elements for one sensor.  The \Kb lines (right panels) present an additional difficulty: because these lines have only one peak, there is a troublesome degeneracy among the three energy-response parameters and the stretch factor $\diff E/\diff P$.  That is, increased instrumental broadening, larger low-energy tails, and larger  $\diff E/\diff P$ all tend to produce similar effects on the spectrum. Thus, \Kb fits in which all eight parameters vary freely tend to fail, or at best to produce an energy-response function that is inconsistent with the \Ka fits: the energy-response does not vary smoothly with photon energy. (The smooth variation of the energy-response is an assumption, but one that is supported by the results of \Ka fits and extensive experience with a range of TES spectrometers.) To rectify this problem, we choose to fix the energy-response parameters using the five \Ka line fits---because the two-peaked fluorescence line breaks these degeneracies---and then to use a linear interpolation of these parameters as a function of energy between the \Ka line energies. With this approach the energy-response parameters are simple and slowly varying functions of photon energy; with only five free parameters, the \Kb fits succeed. The  effective, array-wide energy-response parameters are given in Table~\ref{tab:resolutions}.

\begin{table}
\begin{center}
\begin{tabular}{lrrrr}

Line & Energy (eV) & $\delta E$ (eV)  & \hspace{5mm} $f_\mathrm{tail}$ & $\lambda$ (eV)\\ \hline
Ti K$\alpha_1$ & 4510.90 & 4.18 & 9.5\,\% & 20.0 \\
Ti K$\beta_{1,3}$ & 4931.83 & 4.50 & 8.6\,\% & 16.8 \\
Cr K$\alpha_1$ & 5414.81 & 4.57 & 8.0\,\% & 14.7 \\
Mn K$\alpha_1$ & 5898.80 & 4.77 & 8.1\,\% & 16.3 \\
Cr K$\beta_{1,3}$ & 5946.82 & 4.88 & 8.8\,\% & 15.6 \\
Fe K$\alpha_1$ & 6404.01 & 5.01 & 9.2\,\% & 19.6 \\
Mn K$\beta_{1,3}$ & 6490.59 & 5.19 & 9.6\,\% & 18.4 \\
Co K$\alpha_1$ & 6930.38 & 5.40 & 10.1\,\% & 18.9 \\
Fe K$\beta_{1,3}$ & 7058.18 & 5.46 & 10.6\,\% & 17.9 \\
Co K$\beta_{1,3}$ & 7649.45 & 5.54 & 11.4\,\% & 16.4 \\ \hline
Range & & $\pm0.15$ & $\pm2.0\,\%$ & $\pm2.5$ \\

\end{tabular}
\end{center}
\caption{Parameters of the energy-response function at each of the calibration lines. Line names and energies of the K$\alpha_1$ or K$\beta_{1,3}$ peaks are the first two columns. The response function is determined from fits to each \Ka line; the values at the K$\beta_{1,3}$ lines are interpolated. The energy response is assumed to be a Bortels function; that is, a Gaussian smearing where a fraction $f_\mathrm{tail}$ is also convolved with a low-energy exponential tail of scale length $\lambda$. Column $\delta E$ indicates the full width at half maximum of the Gaussian component. The response is estimated separately for each of 117 active detectors and for each of the three days of measurement. The values here are from the Target~1 observation, and represent a best fit to the combined energy-response function from all detectors. The last line indicates the typical 1$\sigma$ range for parameters across the 117 detectors; this range can also serve as an upper limit to the uncertainty on the parameters.
}
\label{tab:resolutions}
\end{table}

\subsection{Additional calibration lines} \label{sec:cal_additional}

The  K$\alpha_{1,2}$ and K$\beta_{1,3}$ lines of the transition-metal calibrators form the anchors of the sensors' calibration curves. There are numerous additional features present in the measured energy spectra below 4\,keV, however. By including these features, we can explore the calibration curves over a much larger energy range and gain a broader view of the absolute calibration of TES spectrometers. The most prominent low-energy lines are the serendipitous K lines from elements in the target or array environments, including aluminum, silicon, and calcium K$\alpha$. (Although the \Kb lines of Al, Si and Ca and the \Ka lines of Ni and Cu are visible in the combined spectrum from all detectors, none is intense enough in any single-detector spectrum to be useful for calibration.) The line shape from Al \Ka \citep{Lee:2015hh,Wollman:2000kd,Schweppe:1994fc} is used as a proxy for the shape of Si and Ca \Ka as well when fitting to find the peak pulse height. Their true energies are assumed to be those of \citet{Deslattes:2003}: Al: $1486.71\pm0.01$\,eV, Si: $1739.99\pm0.02$\,eV, and Ca: $3691.72\pm0.05$\,eV.

The other calibration features, found between 1.9\,keV and 3.5\,keV, are so-called ``escape peaks.'' Escape peaks are a pair of dim echoes of a higher-energy line at approximately 2.5\,keV below the primary line. They are  caused by the normal absorption of the initial x-ray photon in the bismuth absorbing layer, followed by the re-emission of a bismuth M-line x ray that happens not to be re-captured in the absorber. The bismuth M$\alpha$ and M$\beta$ lines have energies of 2423\,eV and 2525\,eV \citep{Bearden:1967tg}. We include escape peaks in the calibration curve but with a large energy uncertainty of 1\,eV, due to the uncertain M-line energies of bismuth.

Although the uncertainties in both $P$ and $E$ on these additional calibration points far exceed those of the core transition metals, the additional points from 1.5 to 4\,keV are nevertheless useful in the vicinity of 4\,keV to 5.5\,keV, because they better constrain the calibration in that low-energy region of the spectrum.  The unequal intensities of the additional lines mean that while the Al and Si \Ka lines are fit in all cases, the escape peaks and the Ca \Ka line can be fit in  some but not all sensors. Thus the calibration curves all have at least twelve calibration points; most have sixteen to nineteen.

\subsection{Calibration transfer curves} \label{sec:cal_curves}

So far, we have discussed only the data that constrain the $E=f(P)$ curves but not the exact nature of the curves, nor how they are constructed from the constraints. This procedure is critical to the calibration project. Fortunately, we have found that a rather broad range of detailed choices makes little difference to the systematic uncertainties within the interpolated range of energies; the choices do strongly affect the uncertainties when we extrapolate beyond the calibration's anchor points.

Calibration curves must be consistent with the $E$-$P$ measurements to within their uncertainties, but to require an exact match to the data would force an undesirable overfitting on the curves. Some principle is required to choose among the family of possible curves that are all consistent with the measurements, and we use the principle of minimum total curvature. Here, the integrated curvature, $C_f$, of a function $f(x)$ is defined to be the value of $|f''(x)|^2$, integrated over the range from the lowest to the highest measured values of $x$. Other definitions of curvature are possible, but this one leads to a convenient result that the optimal, minimum-curvature function, $f$, is a cubic spline \citep{Green:1994,Reinsch:1967}.

Specifically, we define the penalty function\added{al} $s[f]$ that we wish to minimize as:
\[
s[f] \equiv  \chi^2 + \lambda C_f  = \chi^2 + \lambda \int_{x_\mathrm{min}}^{x_\mathrm{max}}\ |f''(x)|^2\,\mathrm{d}x,
\]
where $\lambda$ is a nonnegative regularization constant, and the quality-of-fit statistic $\chi^2$ is defined as a sum over the $N$ data points:
\[
\chi^2 \equiv \sum_{i=1}^N\ \left [ \frac{E_i - f(P_i)}{\sigma_i}\right]^2.
\]
The equivalent energy uncertainty is
\[
\sigma_i^2 \equiv \delta E_i ^2 + \delta P_i^2\,(\diff E/\diff P)^2.
\]
Here $\delta E_i$ and $\delta P_i$ are the 1$\sigma$ uncertainties on the line energy and the observed pulse height for calibration point number $i$, for $i\in[1,N]$. \citet{Reinsch:1967} shows that \added{for any specific choice of $\lambda$,} the function $f$ that minimizes \added{the functional} $s[f]$ over all  $C^2$ smooth functions is a specific \added{type of} cubic spline, called the \emph{smoothing spline}. It follows also that a smoothing spline minimizes \added{the curvature} $C_f$ subject to a constraint that $\chi^2$ not exceed some fixed limit (such a constraint implies a specific value of $\lambda$).
We choose the limit $\chi^2\le N$ to ensure that the fit be broadly consistent with the data without overfitting.

\removed{Like all cubic splines, smoothing}
\added{Cubic} splines are piecewise cubic-polynomial functions $y=f(x)$, where the third derivatives change discontinuously only at a limited number of locations (``knots''), while the function and its first and second derivatives are continuous everywhere. Smoothing splines \added{specifically} have their knots at each observed $x_i$ but the values of $f(x_i)$ at these knots are not identically the observed values $y_i$.  
Even within the \removed{context} \added{category} of smoothing splines, there are still many possible choices. For one, the spline function $y=f(x)$ need not  identify $y\equiv E$ and $x\equiv P$. It is necessary only that $x$ be some function of the pulse height $P$ and that $y$ be some function of both that is easily solved for energy $E$ given $P$. 
We actually fit using  $y=\log E$ and $x=\log P$.
For another, we use splines satisfying \emph{natural boundary conditions} to remove the two excess degrees of freedom, meaning that $f''(x)$ is constrained to be zero at the lowest and highest knots. We make these choices because they lead to the most successful behavior in the case of extrapolating curves beyond the calibrated range; within that range the choices make no appreciable difference. Extrapolation is never used for extracting metrological data, but it is used repeatedly in the initial stages of building the calibration curve one point at a time, when it can be quite valuable to have curves that extrapolate as accurately as possible. Splines in $\log E$-$\log P$ space have the added advantage that the limit of pulse energy at zero pulse height is naturally zero without an explicit requirement.

It is an active area of research to determine whether some alternative principle for selecting the ``best'' calibration curve is more appropriate than minimization of curvature. Some amount of curvature is intrinsic to the calibration curves of TES sensors, because of the complex physics of the superconducting transition and the non-trivial geometry of a TES. An ideal calibration procedure would somehow penalize curvature due to noise in the data but not the curvature due to TES physics.

\begin{figure}
\begin{center}
\parbox{\linewidth}{
	\includegraphics[width=\linewidth,keepaspectratio]{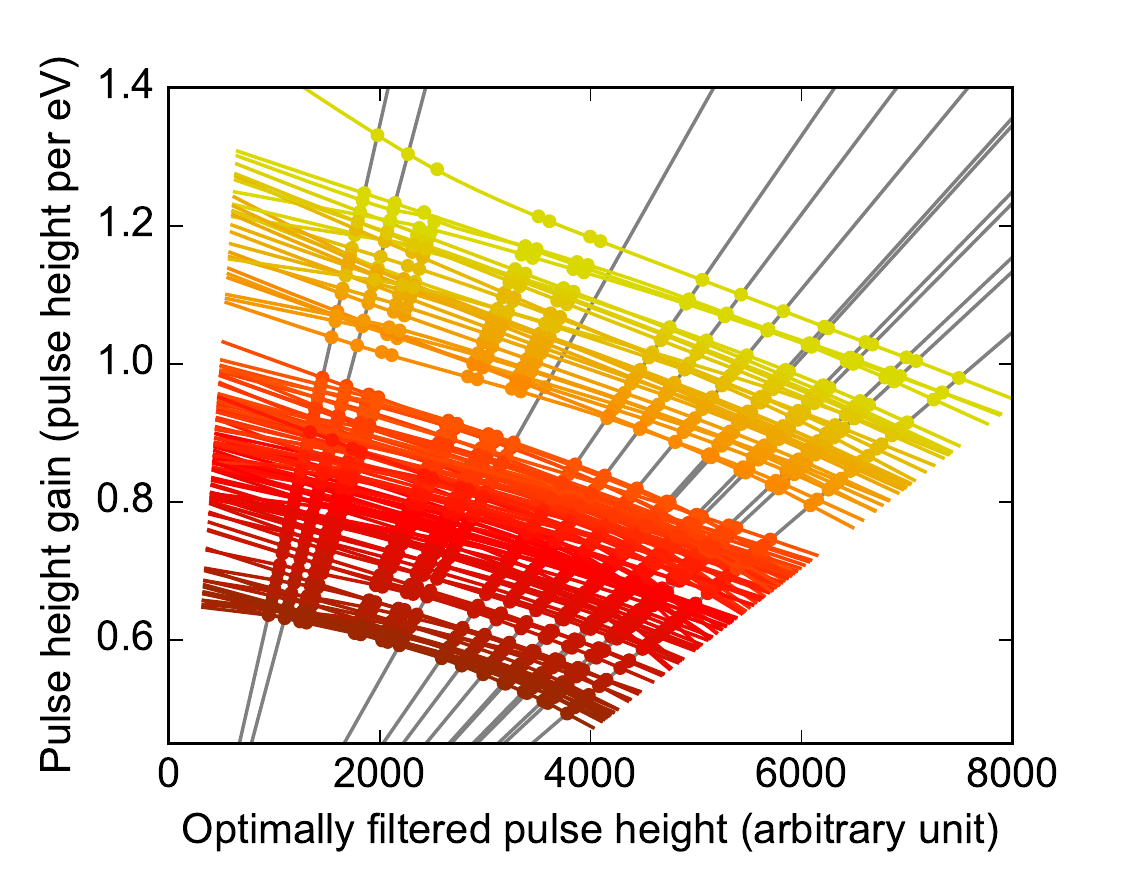}
}
\end{center}
\caption{Gain-calibration curves for 117 sensors. To anchor the curves, we fit the K$\alpha_1$ and K$\beta_{1,3}$ lines of the five calibration elements (as in Figure~\ref{fig:one_channel_calibration}) and find the approximate locations of the faint Al, Si, and Ca \Ka lines and escape peaks (when possible). The optimal choice of interpolating curve is discussed in the text. The gray diagonal lines are lines of constant energy; from left to right they are the \Ka lines of Al, Si, and Ca, and then the 10 \Ka and K$\beta$ lines of the transition metals Ti, Cr, Mn, Fe, and Co. They span the energy range 1487\,eV to 7649\,eV\@. Gray lines are not shown for the escape peaks (2100\,eV to 4500\,eV), as their uncertain absolute calibration reduces their value as calibration points.}
\label{fig:gain_calibration}
\end{figure}

In summary, the calibration curves are of the form $E = \exp[f(\log P)]$ where $f$ is a cubic smoothing spline with natural boundary conditions, which minimizes the total curvature $C_f$ subject to  $\chi^2 \le N$. Examples of calibration points and the approximating curves appear in Figure~\ref{fig:gain_calibration}.

To enable comparison of the linearity of the TES array used in this work to that of other microcalorimeter arrays or even other detector systems, we quantify the nonlinearity in two ways. First, the \emph{gain compression} is the ratio of the gain (pulse height divided by photon energy) at energy $E$ to the gain at 1487\,eV, which is the energy of the lowest calibration point, the Al K$\alpha$ line. The median value of the gain compression is 0.89 at $E=4500$\,eV. At energies of 6000\,eV and 7500\,eV, the median value is 0.84 and 0.78, respectively. A different, locally defined quantity is the \emph{power-law index}. Defined as $(\mathrm{d} \log P/\mathrm{d} \log E)$, it is the index of the power-law curve tangent to the calibration curve at any energy. The median values of the power-law index are 0.82, 0.75, and 0.67 at 4500\,eV, 6000\,eV, and 7500\,eV\@.

\subsection{Assessment of the calibration curves} \label{sec:cal_curve_fidelity}

We assess our calibration curves by a series of leave-one-out cross-validation (CV) tests: a set of new curves is constructed from all but one of the available $P$-$E$ data points for any given sensor, omitting one point at a time. In this way, we can test the fidelity of the curve at each calibration point by comparing the known energy and that estimated by the new curve. From such tests, we find that the calibration curves reproduce the correct value to within approximately 0.1\,eV to 0.3\,eV, provided that the dropped data point is an interior point rather than the highest or lowest. 

Our goal in this study is to estimate the systematic uncertainty inherent to the energy calibration curves. Unfortunately, the CV tests are an imperfect tool for the job. The trouble is that the calibration curves constructed without, for example, the Fe \Ka lines are simply not the same as curves constructed with them. It is not strictly correct to use errors found from the former to predict the errors that will be made when employing the latter. Important (and different) structural information is lost when each different calibration point is omitted and reserved for a validation test. Still, we believe that the CV tests can put a useful figure on the calibration uncertainty, even if it is arguably an upper bound.

\begin{table}
\begin{center}
\begin{tabular}{llrr}
Line & Line &   & \\ 
Name & Energy & Nearest & {\hfill Estimated error\hfill} \\ \hline
Ti K$\alpha_1$ & 4510.90 &  421 & $0.49\pm 0.06$\\
Ti K$\beta_{1,3}$ & 4931.97 &  421 & $0.32\pm 0.04$\\
Cr K$\alpha_1$ & 5414.80 & 484 & $0.32\pm 0.03$\\
Mn K$\alpha_1$ & 5898.80 & 48 & $0.06\pm0.01$\\
Cr K$\beta_{1,3}$ & 5946.82 & 48 & $0.07\pm0.01$\\
Fe K$\alpha_1$ & 6404.01 & 87 & $0.08\pm0.01$\\
Mn K$\beta_{1,3}$ & 6490.18 & 87 & $0.09\pm 0.01$\\
Co K$\alpha_1$ & 6930.38 & 128 & $0.26\pm 0.03$\\
Fe K$\beta_{1,3}$ & 7058.18 & 128 & $0.92\pm 0.05$\\
Co K$\beta_{1,3}$ & 7649.45 & 589 & $20.55\pm 0.81$\\

\end{tabular}
\end{center}
\caption{Calibration curve assessment from 117 sensors on Target~1.  All numbers are in units of eV\@.  \emph{Nearest} is the energy separating the line from its nearest calibration neighbor in either direction. \emph{Estimated eror} is the square root of the mean squared error of the 117 energy values found when the given line is omitted from the construction of the calibration curve, as well as the estimated uncertainty on it. From these values, we construct the heuristic curve (Figure~\ref{fig:cal_bias}) for the  systematic uncertainty due to calibration.  Results for Targets~2 and 3 (not listed) are similar.}
\label{tab:drop_one}
\end{table}

\begin{figure}
\begin{center}
\parbox{\linewidth}{
	\includegraphics[width=\linewidth]{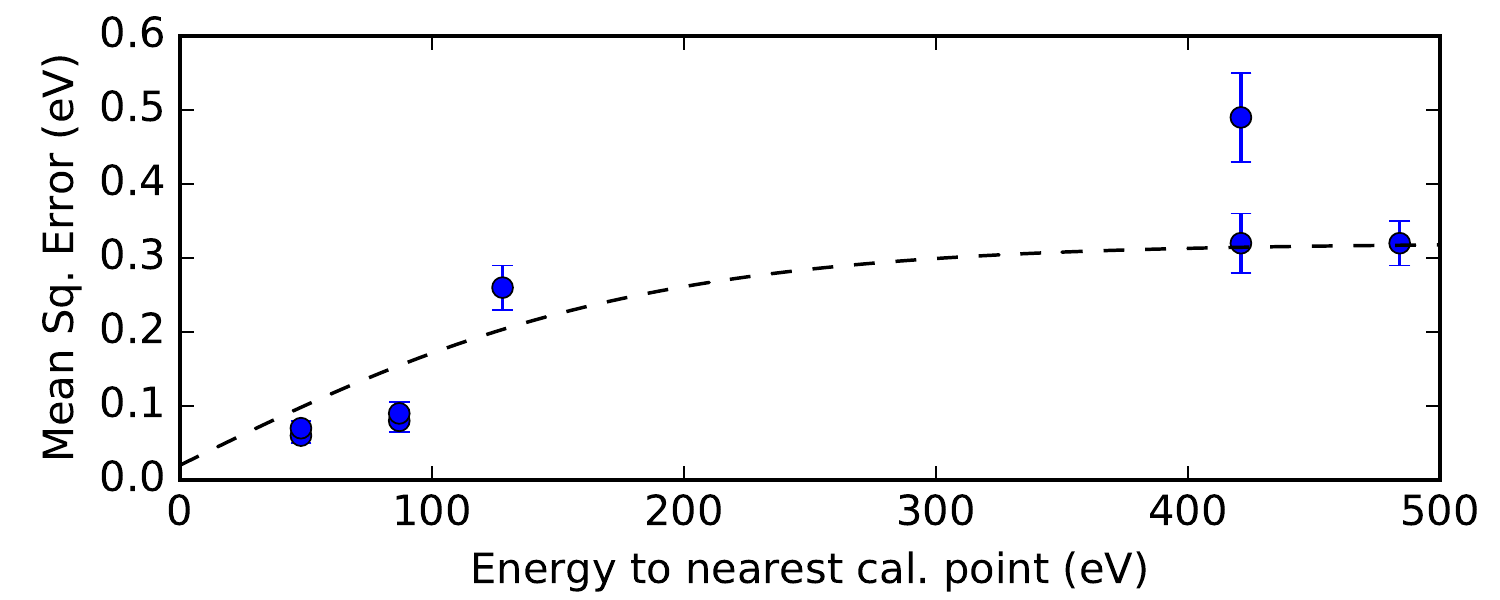}
}
\end{center}
\caption{The square root of the mean-square calibration error, found in leave-one-out cross-validation tests. This error is shown as a function of the energy between the point being tested and the nearest other calibration point (see Table~\ref{tab:drop_one}). The two tests at the high-energy end of the curve are omitted. The dashed line is the heuristic that we employ to approximate the absolute-energy uncertainty due to calibration. Energies above 7\,keV are assigned an additional uncertainty component, as described in the text.}
\label{fig:cal_bias}
\end{figure}

The results are shown in Table~\ref{tab:drop_one}. The estimated error column gives the square root of the mean squared error found in the 117 leave-one-out comparisons.  This estimated error contains contributions from the statistical uncertainty on the calibration data themselves. These contributions, though unwanted, are small enough that we can ignore them safely.
From this table we can see that the systematic energy-assignment error depends weakly on the energy separating an unknown line from the nearest well calibrated line.  We estimate this uncertainty by the heuristic curve shown in Figure~\ref{fig:cal_bias}. The behavior of the calibration curves is possibly less correct above 7000\,eV where the density of anchor points is less; for measurements above this energy, we add a term to the uncertainty $[(E-7000)/650]^2$, which reaches 1\,eV at the top of the calibrated energy range.
The size of the systematic uncertainties due to calibration is a direct result of the nonlinearity of the TES sensors. We anticipate making future measurements with sensors designed to be more nearly linear in the energy range of interest.

\section{Lanthanide Fluorescence Line Results} \label{sec:results}

The prominent L lines of the lanthanide-series elements neodymium, samarium, terbium, and holmium lie between 4.5\,keV and 9\,keV\@. Most of the lines are within the range that is well calibrated by the procedure described in the previous section, that is, up to 7.6\,keV\@.  The emission lines observed in this measurement with sufficient photon counts for further analysis include, from lowest to highest energy, the \Ll, L$\alpha_2$, L$\alpha_1$, L$\beta_1$, L$\beta_3$, L$\beta_2$, and L$\gamma_1$ lines.    Most of these lines feature a single dominant peak, although most also have secondary peaks or some other clear asymmetry. The analysis of these is described in Section~\ref{sec:beta_lines}. The L$\alpha_{2,1}$  doublets require additional analysis steps to find both the peak shape and energy (Section~\ref{sec:alpha_lines}), as the lines in each doublet intrinsically overlap.

With all L lines, we face a problem of model degeneracy similar to the one that arose in fitting the \Kb lines of the transition metals (Section~\ref{sec:cal_transition_metals}): that the parameters of the energy-response function cannot easily be determined because they are degenerate with a model parameter. In the case of the \Kb calibration lines, the degeneracy was with the overall energy calibration ``stretch factor'' $\diff E/\diff P$. For the L lines, the energy scale is already fixed by the calibration procedure. The challenge in the L line analysis is instead that the intrinsic line widths and shapes of the L lines are not well known and are targets of the current study. As with the \Kb lines, we perform all L line fits with the three parameters of the energy-response function (resolution, low-$E$ tail fraction, and tail scale length) fixed at values determined by linear interpolation of values found in the \Ka line fits. 
The uncertainty on the energy-response function leads to systematic uncertainties in the estimated line energies and widths, as discussed in Section~\ref{sec:systematics}.

The combined energy-response function is not exactly a Bortels function when spectra recorded by multiple sensors are combined. Nevertheless, we can estimate the response function for  the combined spectra and fit to it the nearest Bortels function to find an approximate, effective response function of this simple form. For example, the effective energy-response function for the combined data at the Nd L$\beta_1$ line (5720\,eV) has a Gaussian FWHM resolution of 4.92\,eV and a 9.4\,\% tail fraction with a scale length of 17\,eV\@. The error introduced by this effective response function is discussed in Section~\ref{sec:systematics}  as item~\ref{syserr:bortels} and is found to be small.

\subsection{Singlet emission lines} \label{sec:beta_lines}

\newcommand{\Nc}{\ensuremath{N_\mathrm{c}}}

The L$\beta_3$ lines and some \Ll\ lines are symmetric to within our ability to resolve their shape and are well described by a single Lorentzian. The other singlet lines (L$\beta_1$, L$\beta_2$, and L$\gamma_1$) have a more complex shape, owing to the presence of satellite lines that arise from processes involving more than one electron; in these cases, two or more Lorentzian components are required to capture their full line shapes.  The line shape models are fully described by three parameters per Lorentzian component: the overall intensity, the center energy and the Lorentzian width.

\begin{figure}
\begin{center}
\parbox{\linewidth}{

	\includegraphics[width=\linewidth,keepaspectratio]{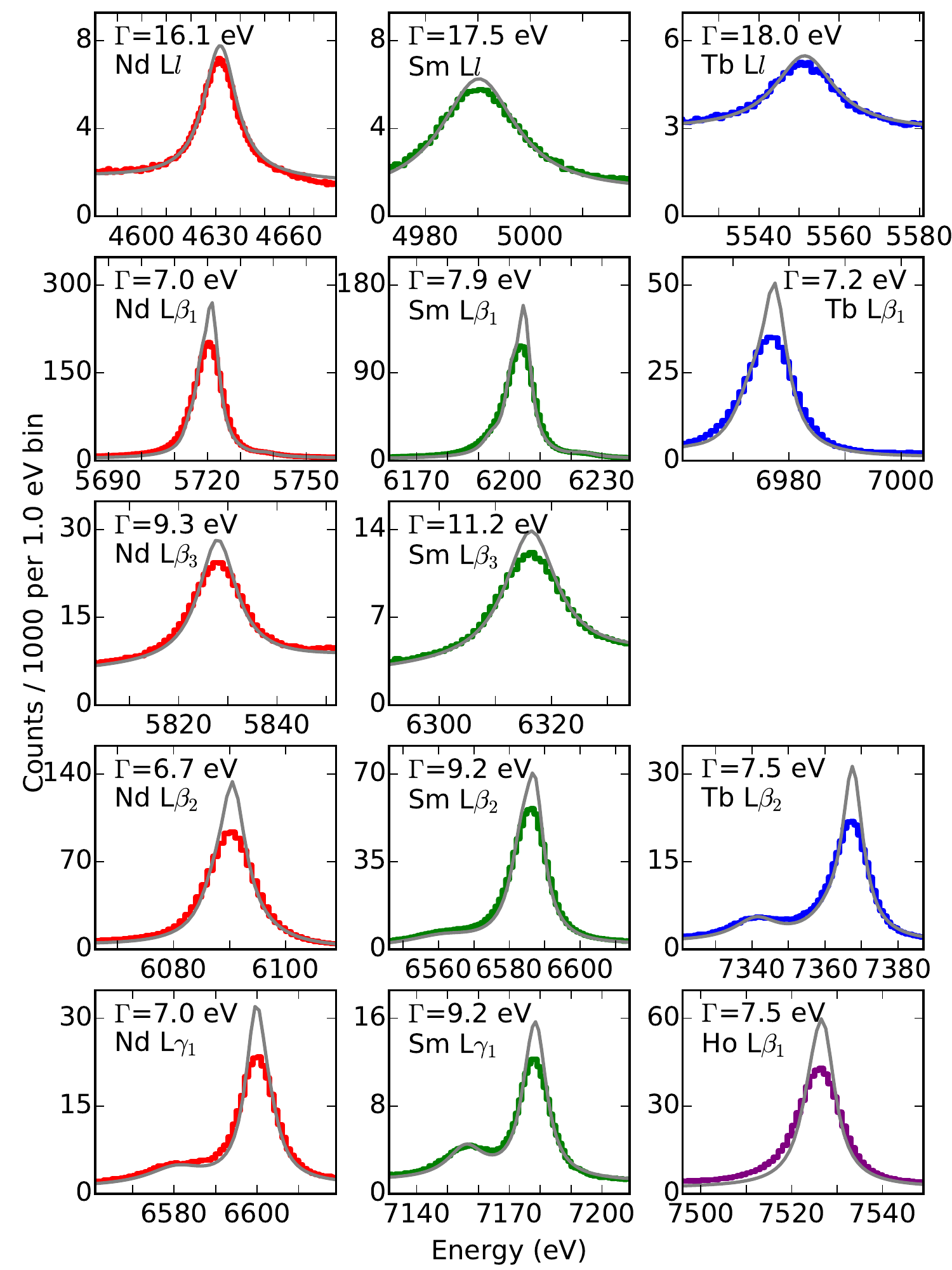}
}
\end{center}
\caption{The fourteen lanthanide L-line singles measured in this study, including lines \Ll, L$\beta_1$, L$\beta_3$, L$\beta_2$, and L$\gamma_1$. Data from all sensors are combined. The Tb L$\beta_3$ line is visible in the full spectrum but is not bright enough for us to assess its width. The Tb L$\gamma_1$ line and most lines of Ho are inaccessible; they are either masked by brighter calibration lines or lie at energies above 7.6\,keV, the upper limit of the well calibrated energy range. The peak signal level ranges from 1500 (Tb \Ll) to 200,000 (Nd L$\beta_1$) counts above background per 1\,eV bin.   Each histogram shows our measured spectrum; the smooth gray curve shows the best fit model (including background counts) with the instrumental broadening set to zero (that is, the resolution and tail effects are removed from the model).
The value $\Gamma$ noted in each panel is the FWHM of the line's best-fit model as a sum of one or more Lorentzian components (the observed FWHM is wider, as it includes the effect of instrumental broadening). Uncertainties on $\Gamma$ are given in Table~\ref{tab:Lbeta_widths}.
\label{fig:Lbeta_fits} }
\end{figure}

To estimate the line shapes, we combine the spectra from all calibrated TESs, shown in Figure~\ref{fig:Lbeta_fits}. We then perform one fit to the combined data for the widths, amplitudes, and center energies of each of $\Nc$ Lorentzian components. We also allow for a background level that varies linearly in the energy, for a total of $(2+3\Nc)$ variables in each \Nc-component fit. Each fit is made to spectra with two bins per eV over an energy range of 45\,eV to 90\,eV in width (depending on the energy spanned by the visible line structure). Most fits are centered near the expected line peak, but in a few cases the limits are shifted off-center to avoid contamination from nearby emission lines of the calibration elements (such as the Cr \Kb line at 4932\,eV and the Fe \Ka line at 6404\,eV).

The L$\beta_1$, L$\beta_2$, and L$\gamma_1$ lines are found to be inconsistent with a single-Lorentzian model; that is, they require $\Nc>1$. We have therefore repeated the fits, adding Lorentzian components until the reduced $\chi^2$ of the fit is consistent with noise or until additional terms no longer contain at least 2\,\% of the total line intensity. This procedure requires as many as five Lorentzians in the current data, as indicated in Table~\ref{tab:Lbeta_widths} by the column $\Nc$. The Nd and Sm L$\beta_1$ lines, in particular, are poorly fit unless the model includes a small, wide component centered some 15\,eV to 18\,eV above the main peak. These components, at $5738\pm2$\,eV and  $6224\pm2$\,eV, might constitute an observation of the L$_2$M$_5$ lines, though these are predicted by theory to be at $5744.6\pm1.1$\,eV and $6232.6\pm1.2$\,eV \citep{Deslattes:2003}. No comparable suggestion of L$_2$M$_5$ lines is apparent above the Tb or Ho L$\beta_1$ lines.
The Nd and Sm L$\beta_1$ lines are also best considered as L$\beta_{1,4}$ blends, given the near-alignment of the L$\beta_4$ (L$_1$M$_2$) and L$\beta_1$ transition energies of these elements; this blending is another reason to expect $\Nc>1$ for these two cases.

For all lines, we define the line energy as the energy at the maximum of the multi-Lorentzian model, which is numerically determined in the limit of zero instrumental broadening. Multi-component models introduce a systematic uncertainty on the line energies and widths that follows from the uncertain details of the model. This effect is discussed in Section~\ref{sec:systematics} as item \ref{syserr:lineshape}.  Alternative quantities that capture the energy of a line complex were also considered, including the line centroid (a weighted mean of the center energies of the \Nc\ separate Lorentzian components) and the median (the 50 percentile point of the model's cumulative density function). The centroid and median of the models were in some cases more robust as statistical estimators than the peak energy, but not universally so. Furthermore, among these possible energy descriptions, only the models' peak energy is directly comparable to the reference data.

We summarize the line shapes by our estimates of the line full-width at half maximum, $\Gamma$, which appear in Figure~\ref{fig:Lbeta_fits} and in Table~\ref{tab:Lbeta_widths}. The value of $\Gamma$ is computed by numerically locating the peak's maximum and half-maximum points. While $\Gamma$ is an incomplete summary of complex line shapes, we consider it to be a useful first step towards fuller parameterization of the shapes in future work. \added{The value is compared with theoretical estimates (here, the sum of initial-state and final-state widths) from~\citet{Campbell:2001}. For lines with predicted widths between 4\,eV and 6\,eV, the measurements are systematically larger than the theory. This discrepancy could reflect incomplete deconvolution of the TES energy-response function, but we attribute it mainly to the fact that these complex emission lines are the result of processes beyond the simple transition of a single electron from a single initial state to a single final state that the theoretical value assumes.}

\begin{table}
\begin{center}
\begin{tabular}{llrrr}
\multicolumn{2}{c}{Line name} & \\ 
Siegbahn & IUPAC  & $N_\mathrm{c}$ & $\Gamma \pm$ stat. $\pm$ syst. & \added{Theory} \\ \hline
Nd \Ll & L$_3$M$_1$ & 2 & $16.15 \pm 0.27 \pm 0.29$ & 15.4 \\
Nd L$\beta_1$ & L$_2$M$_4$ &4 &$7.03 \pm 0.13 \pm 0.62$ & 4.3\\
Nd L$\beta_3$ & L$_1$M$_3$ &1 &$9.33 \pm 0.09 \pm 0.40$ & 10.2\\
Nd L$\beta_2$ & L$_3$N$_5$ &3 & $6.72 \pm 0.12 \pm 0.55$ & 4.2 \\
Nd L$\gamma_1$ & L$_2$N$_4$ &3 & $6.98 \pm 0.20 \pm 0.43$ & 4.6\\
Sm \Ll & L$_3$M$_1$ & 1 & $17.46 \pm 0.35 \pm 0.66$ & 15.9\\
Sm L$\beta_1$ & L$_2$M$_4$ &4 & $7.89 \pm 0.33 \pm 0.53$ & 4.6 \\
Sm L$\beta_3$ & L$_1$M$_3$  &1& $11.21 \pm 0.15 \pm 0.35$  & 11.4\\
Sm L$\beta_2$ & L$_3$N$_5$ &5 & $9.20 \pm 0.08 \pm 0.42$  & 5.0\\
Sm L$\gamma_1$ & L$_2$N$_4$ &3 & $9.17 \pm 0.20 \pm 0.44$ & 5.5\\
Tb \Ll & L$_3$M$_1$ &1 & $18.02 \pm 0.54 \pm 3.01$ & 16.9\\
Tb L$\beta_1$ & L$_2$M$_4$ &3 & $7.16 \pm 0.20 \pm 0.63$ & 4.9 \\
Tb L$\beta_2$ & L$_3$N$_5$ &3 & $7.46 \pm 0.07 \pm 0.39$ & 6.2\\
Ho L$\beta_1$ & L$_2$M$_4$ &2 & $7.46 \pm 0.21 \pm 0.48$  & 5.2\\
\end{tabular}
\end{center}
\caption {Estimates of line widths $\Gamma$ (full width at half maximum, in eV) for the lines that appear in Figure~\ref{fig:Lbeta_fits}, in a model with $N_\mathrm{c}$ Lorentzian components. The statistical uncertainties are the standard deviations of the fit value of $\Gamma$ produced by simulating the data distribution and re-fitting 500 times. The systematic uncertainties (see Section~\ref{sec:systematics}) are generally dominated by the uncertainty in the energy resolution or in the line shape model. \added{The last column gives theoretical estimates of line widths~\cite{Campbell:2001}.}
}
\label{tab:Lbeta_widths}
\end{table}

\begin{figure}
\begin{center}
\parbox{\linewidth}{
	\includegraphics[width=\linewidth,keepaspectratio]{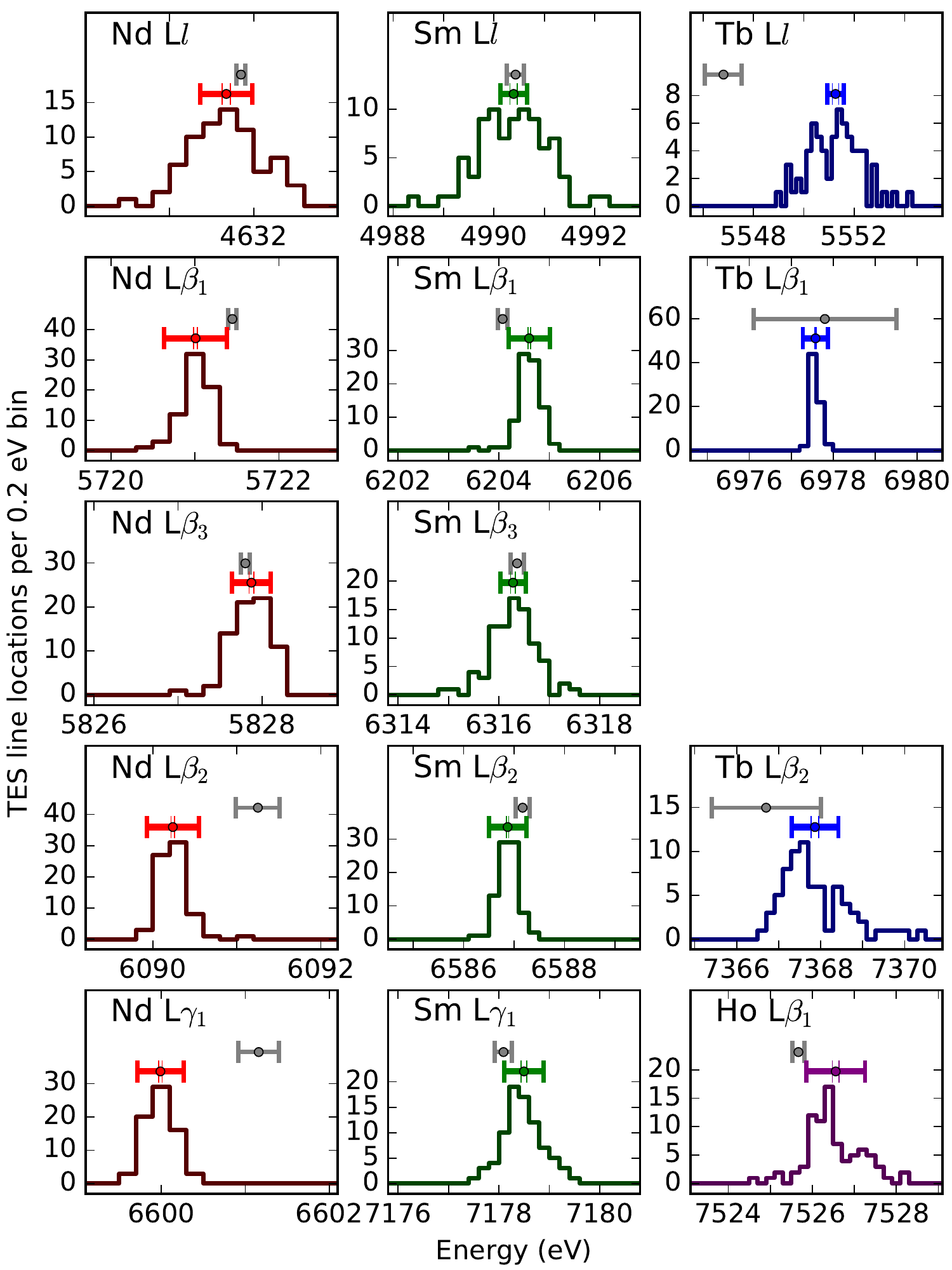}
}
\end{center}
\caption{The line-energy estimates for fourteen lanthanide L-line singlets. The histograms reflect the distribution of the line-energy fit made one TES detector at a time. The error bars above this show the central estimates and 1$\sigma$ uncertainties on the peak energy: \citet{Deslattes:2003} (top bar, gray), and this work (lower bar). The latter error bars at their full extent show the uncertainties including systematics; the inner, thinner vertical ticks reflect only the standard error on the mean of the measured values and thus exclude our estimate of the correlated systematic uncertainty. }
\label{fig:Lbeta_locations}
\end{figure}

Once the line shape is determined from the all-TES spectrum, we fit each separate TES spectrum---now holding the line shape fixed---to get a separate estimate of the line energy from each sensor. Unlike the fit for the line shape, successful fits for the peak location can be performed even with the limited statistics available from a single sensor. Figure~\ref{fig:Lbeta_locations} shows the distribution of these estimates for each singlet line. Table~\ref{tab:Lbeta_locations} gives the mean energy of the peaks and the standard error of the mean as the statistical uncertainty on it. Robust estimators of location and scale were also considered (e.g., \citet{Rousseeuw:1993}), to minimize our sensitivity to a small number of outliers. These estimators differ from the mean and standard error by less than 0.1\,eV in all cases, usually by less than 0.02\,eV; both are well within the systematic uncertainties on the measurements. 

The reported peaks in Table~\ref{tab:Lbeta_locations} are the peak energies of the combined \Nc\ Lorentz functions after correction for our sensors' energy-response function. The observed peak in the microcalorimeter data is at slightly lower energies than the reported peaks, because of the asymmetric energy-response function with its tail to low energies. The size of this effect depends on the line shape and is typically a few tenths of an eV (see Figure~\ref{fig:peak_smear_all_lines}).

Although we could re-estimate the line widths and shapes by shifting the individual sensors' spectra to align the separate peaks and iterating the entire procedure, in practice, we find that in no case below 7.6\,keV does a second iteration change the line width by more than a small fraction of the estimated uncertainties.

\begin{table}
\begin{tabular}{lrr}
Line & This work $\pm$ stat $\pm$ syst & Reference $\pm$ err \\ \hline
Nd L$l$ & 4631.67 $\pm$ 0.05 $\pm$ 0.29 & 4631.85 $\pm$ 0.05\\
Nd L$\beta_1$ & 5721.01 $\pm$ 0.02 $\pm$ 0.37 & 5721.45 $\pm$ 0.05\\
Nd L$\beta_3$ & 5827.87 $\pm$ 0.03 $\pm$ 0.23 & 5827.80 $\pm$ 0.05\\
Nd L$\beta_2$ & 6090.24 $\pm$ 0.02 $\pm$ 0.31 & 6091.25 $\pm$ 0.26\\
Nd L$\gamma_1$ & 6599.99 $\pm$ 0.02 $\pm$ 0.28 & 6601.16 $\pm$ 0.24\\
Sm L$l$ & 4990.39 $\pm$ 0.07 $\pm$ 0.26 & 4990.43 $\pm$ 0.17\\
Sm L$\beta_1$ & 6204.60 $\pm$ 0.03 $\pm$ 0.37 & 6204.07 $\pm$ 0.09\\
Sm L$\beta_3$ & 6316.28 $\pm$ 0.05 $\pm$ 0.25 & 6316.36 $\pm$ 0.13\\
Sm L$\beta_2$ & 6586.87 $\pm$ 0.02 $\pm$ 0.37 & 6587.17 $\pm$ 0.14\\
Sm L$\gamma_1$ & 7178.50 $\pm$ 0.05 $\pm$ 0.36 & 7178.09 $\pm$ 0.17\\
Tb L$l$ & 5551.26 $\pm$ 0.12 $\pm$ 0.30 & 5546.81 $\pm$ 0.73\\
Tb L$\beta_1$ & 6977.58 $\pm$ 0.01 $\pm$ 0.30 & 6977.80 $\pm$ 1.70\\
Tb L$\beta_2$ & 7367.86 $\pm$ 0.09 $\pm$ 0.55 & 7366.70 $\pm$ 1.30\\
Ho L$\beta_1$ & 7526.56 $\pm$ 0.08 $\pm$ 0.69 & 7525.67 $\pm$ 0.15\\
\end{tabular}

\caption{Estimates of the line energies for the fourteen L-line singlets. \emph{This work} indicates our measured line peak energy, based on fits to the single-TES spectra as one or multiple Lorentz functions, smeared by our known energy-response function for that TES at the given energy.  The statistical uncertainty column (\emph{stat}) gives the spread of the measured values divided by the square root of the number of measurements, while the systematic (\emph{syst}) is described in Section~\ref{sec:systematics}. The latter is generally dominated by the calibration curve interpolation uncertainty, which we learn from the results presented in Table~\ref{tab:drop_one}, and by the error introduced by learning the line shape from our data. The \emph{Reference} and \emph{err} columns are the central value and 1$\sigma$ uncertainties reported in the reference data, \citet{Deslattes:2003}.}
\label{tab:Lbeta_locations}
\end{table} 

\subsection{\La  doublets} \label{sec:alpha_lines}

\begin{figure}
\begin{center}
\parbox{\linewidth}{
	\includegraphics[width=\linewidth,keepaspectratio]{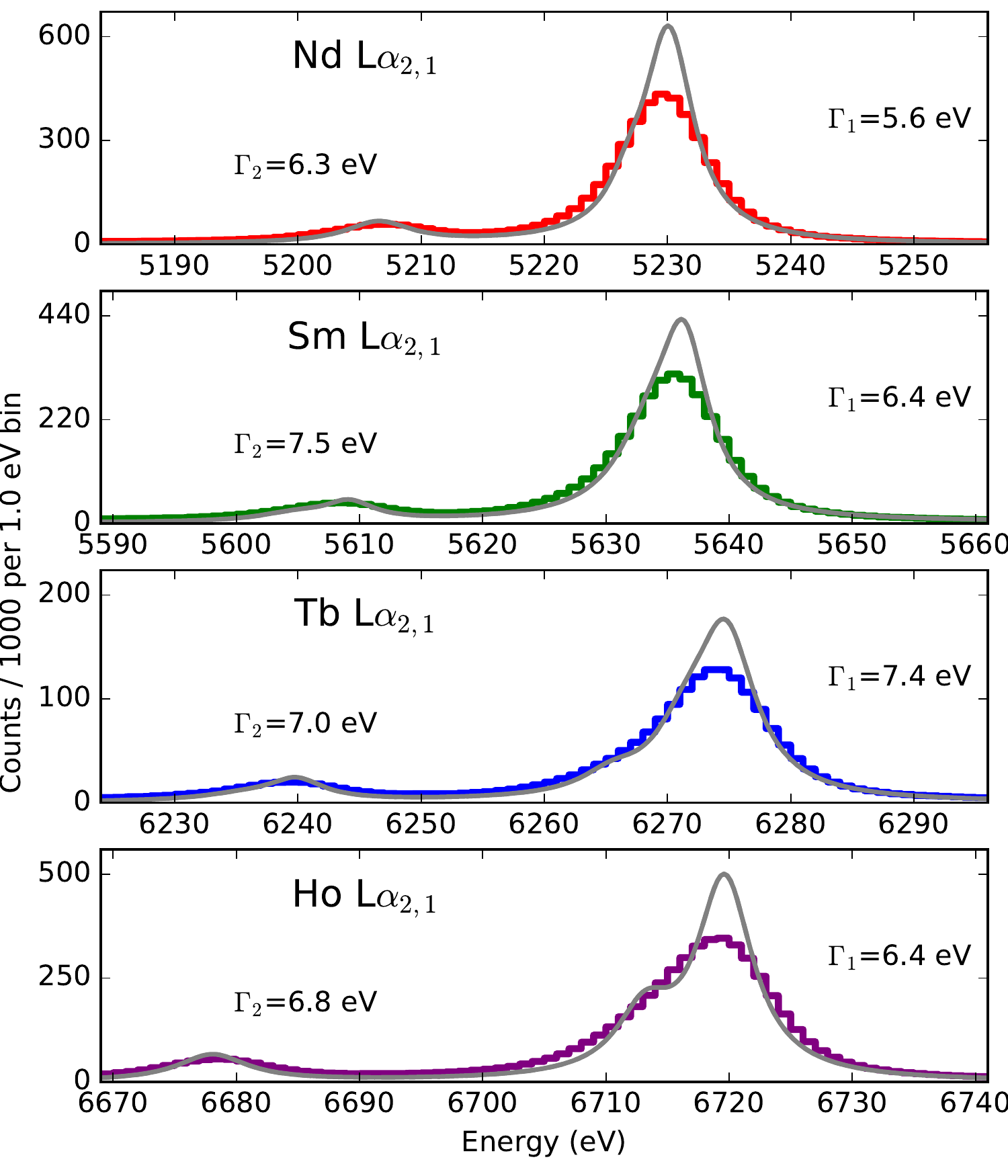}
}
\end{center}
\caption{The line shapes of lanthanide \La line doublets, for elements Nd, Sm, Tb, and Ho. The measurements are shown as histograms. The peak signal level ranges from 100,000 to 400,000 counts per 1\,eV bin.   
Spectra are all 72\,eV wide and centered between the L$\alpha_2$ and L$\alpha_1$ lines. The values $\Gamma$ indicated in each plot are each peak's FWHM, after a model is fit and background and effects of instrumental broadening are removed. The resulting estimate of the intrinsic line shape is shown as a smooth, gray curve.
\label{fig:Lalpha_fits} }
\end{figure}

The lanthanide elements' \La lines are similar to the L$\beta_1$ and other asymmetric lines in that the L$\alpha_2$ and L$\alpha_1$ lines can only be modeled as a sum of multiple Lorentzian components. As with the singlet lines, we first determine the line \emph{shape} using the all-TES combined energy spectrum, because only the combined spectrum has enough x-ray counts for the task.
The shape fit again assumes the effective Bortels response function found from a fit to the combined response function. The \La line shapes require $\Nc=3$ Lorentzian components for Nd and Ho, and $\Nc=5$ for Sm and Tb. Additional peaks beyond the first two allow an asymmetry in one or both L$\alpha$ lines. Figure~\ref{fig:Lalpha_fits} shows the measurement and the full \Nc-component model without background or resolution effects. Line widths are given in Table~\ref{tab:Lalpha_widths} and indicate the numerical full-width at half maximum of each peak in the zero-background, perfect-resolution model.
\added{The value is compared with theoretical estimates from~\citet{Campbell:2001}. The measurements are systematically larger than the theory, as in the previous section, and we again attribute the discrepancy to the rich array of multi-electron transitions not accounted for in the theoretical values.}

\begin{figure}[t]
\begin{center}
\parbox{\linewidth}{
	\includegraphics[width=\linewidth,keepaspectratio]{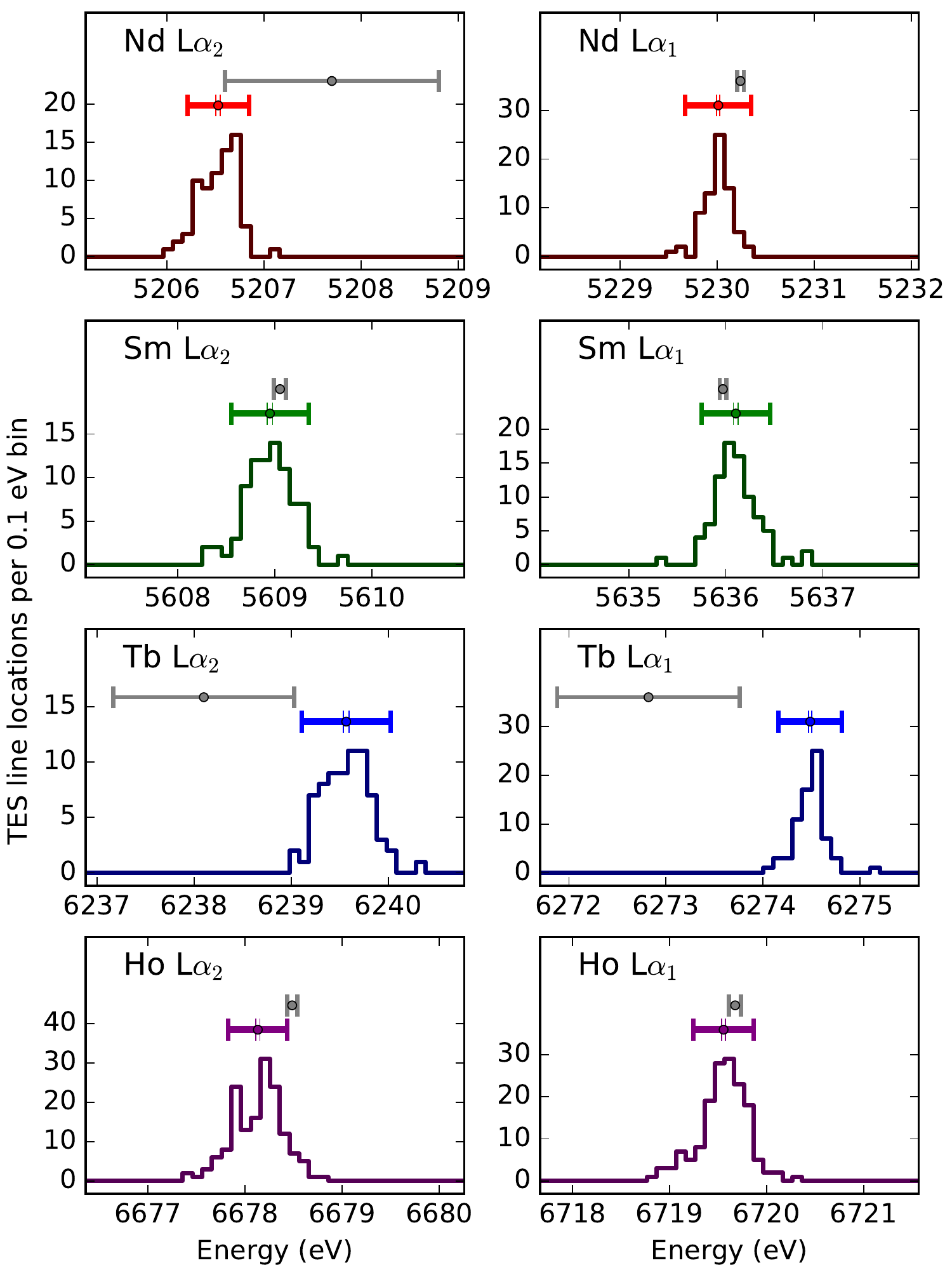}
}
\end{center}
\caption{The line energy estimates for eight L$\alpha$ lines. The histograms reflect the distribution of the line peak energy from fits made one TES detector at a time. The error bars above this show the central estimates and 1$\sigma$ uncertainties on the peak energy: \citet{Deslattes:2003} (top bar, gray), and this work (lower bar). As in Figure~\ref{tab:Lbeta_locations}, the lower error bars show both the full uncertainties including systematics (full bar) and the statistical uncertainties only (narrow ticks). }
\label{fig:Lalpha_locations}
\end{figure}

The peak \emph{energies} of the \La lines are determined by performing two separate fits to the spectrum recorded by each TES. In these fits, the same composite line shape (the shape of the entire doublet) is held fixed while an overall energy shift is allowed in each. One fit uses only the  L$\alpha_2$ portion of the measured spectrum to estimate that line's peak energy, and the other uses only the L$\alpha_1$ portion; in each, a linear function is added to represent the background count rate. Both fits compare the energy-limited range of the measured data to the model of the complete doublet.  The best-fit shifts in each case yield our estimators for the peak energy of the \La lines. Figure~\ref{fig:Lalpha_locations} shows the distribution of the peak energy estimates, as well as the range of values consistent with previous work \citep{Deslattes:2003} and then with this work, either excluding or including the effects of systematic uncertainty in the calibration. The wider, inclusive uncertainties in Figures~\ref{fig:Lbeta_locations} and \ref{fig:Lalpha_locations} indicate our best current estimate of the peak energies. The exclusive uncertainties give a sense of what might be achieved with this technique as the sensors are made more linear and with higher resolution, and as the density of well-known calibration lines is increased. Prospects for improvements along these lines are explored in Section~\ref{sec:prospects}. The estimated line energies appear in Table~\ref{tab:Lalpha_locations}.

\newcommand{\raiseme}[1]{\raisebox{1.5ex}[0pt]{#1\hspace{6pt}}} 
\begin{table}
\begin{center}
\begin{tabular}{llrrr}
\multicolumn{2}{c}{Line name} & \\ 
Siegbahn & IUPAC  & \Nc & $\Gamma \pm$ stat. $\pm$ syst. & \added{Theory} \\ \hline
Nd L$\alpha_2$ & L$_3$M$_4$ && $6.33 \pm 0.06 \pm 0.46$ & 4.1 \\
Nd L$\alpha_1$ & L$_3$M$_5$ &\raiseme{3} & $5.57 \pm 0.04 \pm 0.52$ & 4.1\\
Sm L$\alpha_2$ & L$_3$M$_4$ && $7.55 \pm 0.05 \pm 0.73$ & 4.4 \\
Sm L$\alpha_1$ & L$_3$M$_5$ &\raiseme{5} & $6.44 \pm 0.03 \pm 0.41$ & 4.4\\
Tb L$\alpha_2$ & L$_3$M$_4$ && $6.95 \pm 0.12 \pm 0.65$ & 4.8\\
Tb L$\alpha_1$ & L$_3$M$_5$ &\raiseme{5} & $7.42 \pm 0.08 \pm 0.50$ & 4.8\\
Ho L$\alpha_2$ & L$_3$M$_4$ && $6.85 \pm 0.05 \pm 0.44$ & 5.1\\
Ho L$\alpha_1$ & L$_3$M$_5$ &\raiseme{3} & $6.38 \pm 0.03 \pm 0.46$ & 5.1\\
\end{tabular}
\end{center}
\caption {Estimates of the line full-width at half-maximum, in eV, for the lines shown in Figure~\ref{fig:Lalpha_fits}. The value $\Gamma$ and the statistical and systematic errors are as described in the caption for Table~\ref{tab:Lbeta_widths}. In the case of \La lines, a single line shape model with \Nc\ Lorentzian components describes both peaks that make up the \La doublet.  \added{The last column gives theoretical estimates of line widths~\cite{Campbell:2001}.}
} \label{tab:Lalpha_widths}
\end{table}

\begin{table}
\begin{tabular}{lrrrrr}
Line & This work $\pm$ stat $\pm$ syst & Reference $\pm$ err \\ \hline
Nd L$\alpha_2$ & 5206.53 $\pm$ 0.02 $\pm$ 0.32 & 5207.70 $\pm$ 1.10\\
Nd L$\alpha_1$ & 5230.01 $\pm$ 0.02 $\pm$ 0.33 & 5230.24 $\pm$ 0.04\\
Sm L$\alpha_2$ & 5608.95 $\pm$ 0.03 $\pm$ 0.40 & 5609.05 $\pm$ 0.06\\
Sm L$\alpha_1$ & 5636.10 $\pm$ 0.03 $\pm$ 0.35 & 5635.97 $\pm$ 0.03\\
Tb L$\alpha_2$ & 6239.57 $\pm$ 0.03 $\pm$ 0.40 & 6238.10 $\pm$ 0.93\\
Tb L$\alpha_1$ & 6274.49 $\pm$ 0.02 $\pm$ 0.31 & 6272.82 $\pm$ 0.94\\
Ho L$\alpha_2$ & 6678.13 $\pm$ 0.02 $\pm$ 0.30 & 6678.48 $\pm$ 0.05\\
Ho L$\alpha_1$ & 6719.56 $\pm$ 0.02 $\pm$ 0.31 & 6719.68 $\pm$ 0.06\\
\end{tabular}

\caption{Estimates of the line energies for the eight L$\alpha$ lines. \emph{This work} indicates our measured line peak energy, based on numerical evaluation of the best-fit multi-Lorentzian model with ideal energy response and averaged over all TESs.  The next column (\emph{stat}) gives the spread of the measured values divided by the square root of the number of measurements, while the systematic (\emph{syst}) is described in Section~\ref{sec:systematics}.  The \emph{Reference} and \emph{err} columns are the central value and 1$\sigma$ uncertainties reported 
in \citet{Deslattes:2003}.
}
\label{tab:Lalpha_locations}
\end{table}

\subsection{Systematic uncertainties on line energies and widths} \label{sec:systematics}

\begin{table*}
\begin{center}
\setlength{\tabcolsep}{1.5ex}
\begin{tabular}{lll}
Source of systematic uncertainty & syst($E_\mathrm{pk}$) & syst($\Gamma$) \\ \hline
1. Free vs fixed low-E model & 0.003 & 0.005 \\
2. Energy-dependent QE & 0.005 & 0.01 \\
3. Low-E tails of nearby lines  & 0.01 & $ 0.002$ \\
4. Histogram bin density & 0.01 & 0.02 \\
5. Uncertain low-E tail model & 0.02 & 0.05 \\
6. Composite $E$-response model & 0.03 & 0.03 \\
7. Uncertain energy resolution & 0.03 to 0.15 & 0.2 to 1.0 \\
8. Energy range used in fit & up to 0.15 & 0.1 to 0.5 \\
9. Calibration interpolation  & 0.1 to 0.6 & 0.01 to 0.05 \\
10. Fit model for L line shapes & 0.2 & 0.01 to 0.8 \\
\end{tabular}
\end{center}
\caption{Typical systematic uncertainties on the peak energies ($E_\mathrm{pk}$) and on the full width at half maximum ($\Gamma$) for the 22 L lines. All values are in eV\@. The causes are described more fully in Section~\ref{sec:systematics}. The overall shift due to the interpolation method (\#\ref{syserr:calibration})---which increases with energy difference between the peak and the nearest calibration anchor---and the unknown exact line shape  (\#\ref{syserr:lineshape}) are the two dominant systematics on $E_\mathrm{pk}$. The dominant uncertainty on the estimates of $\Gamma$ results from uncertainty on the exact energy resolution (\#\ref{syserr:resolution}), though a few values of $\Gamma$ also have a large contribution from the uncertain line shape model (\#\ref{syserr:lineshape}).
} \label{tab:systematics}
\end{table*}

The uncertainties on the peak energy and width of the various emission lines are in most cases dominated not by statistical uncertainties (that is, uncertainties that should improve as more data are collected) but by systematic uncertainties. We have studied the effects of many different analysis choices and other potential sources of overall uncertainties, summarizing the results in the systematic uncertainty budget, Table~\ref{tab:systematics}. Some entries are extremely small and are listed only for the sake of completeness; others are important components of the uncertainties on either the peak energies or their widths, or both.

The entries are listed approximately in order of increasing effect on the peak energy. A description of each follows.
\begin{enumerate}
\item The parameters of the low-energy tail are fit only from the \Ka lines of the 3d transition metals and interpolated to all other energies. If they were instead fit to the data at each lanthanide L line, it would have an effect of only a few meV on peak energy or width estimates. (It would also increase systematic uncertainties.)
%
\item\label{systematic:QE} The quantum efficiency of the system depends on photon energy. Fortunately, the transmission of the filters grows with increasing energy in a manner that largely cancels out the reduced absorption of the detectors, and the effect across the width of narrow fluorescence lines is a few meV at most.
\item\label{syserr:tails} The small low-energy tails of the next line or lines at energies just above the region of interest 
are handled by allowing a linearly varying background level in each fit. The effect of not modeling these tails more explicitly is found to be unimportant.
\item The peak energy and width can vary by 0.01\,eV and 0.02\,eV, respectively, depending on the bin spacing of the spectra.
\item When the parameters of the low-energy-tail component of the energy-response function are varied by their systematic uncertainties (typically $\pm1.5\,\%$ on $f_\mathrm{tail}$ and $\pm2$\,eV on its scale length), the peak energies and widths change by a few times 0.01\,eV\@.
\item \label{syserr:bortels} It is expedient to model the combination of the energy-response functions of the $\sim100$ sensors---each with a distinct Gaussian resolution and low-energy tail---as a single Gaussian plus a tail, with parameters chosen appropriately for the energy in question. This approximation affects the peak energies and widths by no more than 0.03\,eV, determined by comparisons between models that were fit with and without this simplification.
\item \label{syserr:resolution} The uncertainty on the energy resolution has a small effect on the peak energy but is the dominant source of uncertainty on our estimates of the width. We vary (the Gaussian component of) the resolution by $\pm 0.08$\,eV (which is the typical 1$\sigma$ uncertainty) and repeatedly fit simulated spectra. The resolution errors typically add uncertainties of 0.1\,eV to the peak energy and 0.5\,eV to the width.
\item If we vary (by a few eV) the exact range of energies used when we fit for the line shape, the choice can affect the peak energies by approximately 0.05\,eV to 0.15\,eV and widths by up to a few tenths of an eV\@.  The effect is larger than item \ref{syserr:tails}, which accounts for the incomplete model of the background. This suggests that the chosen energy range affects the peak energy indirectly, as fits to the line shapes change slightly when the energy range is varied.
\item \label{syserr:calibration} Uncertainty in absolute energies arises from the interpolation procedure for calibration curves (the topic of Section~\ref{sec:cal_curve_fidelity} and Table~\ref{tab:drop_one}). It can be the dominant systematic effect on the peak energies, though the amount of uncertainty ranges from 0.10\,eV for some fluorescence lines that lie within approximately 100\,eV of the nearest calibration anchor point, to as much as 0.5\,eV to 0.6\,eV for lines with energies that exceed 7300\,eV\@. A typical value is 0.25\,eV\@. Uncertainty in the exact calibration curve has only a minor effect on our estimates of peak widths.
\item \label{syserr:lineshape} A few singlet lines are modeled as a single Lorentzian, but \added{the complexity of atomic transitions means that} most lines have to be treated as a sum of multiple Lorentzians. This model cannot be exact in all details, given the spectrometer's energy resolution of 5\,eV to 6\,eV\@. Simulated data from multiple-Lorentz models similar to the lanthanide L lines show that the susceptibility of a line to this systematic error depends on its detailed shape. We find that fits of our data to our own shape model can disagree by $\pm0.2$\,eV with fits of the same data to models provided by wavelength-dispersive instruments. Such models are available for the transition K$\beta_{1,3}$ lines and certain L lines. We cannot say whether this discrepancy is attributable to deficiencies in our work or in earlier work, so we adopt the conservative uncertainty of 0.2\,eV on line energies due to this effect.
\end{enumerate}
We have summarized these effects only in general terms here and in Table~\ref{tab:systematics}, but we compute a specific value of each item for each fluorescence line. The quadrature sum of all effects is given in Tables~\ref{tab:Lbeta_widths}, \ref{tab:Lbeta_locations}, \ref{tab:Lalpha_widths}, and \ref{tab:Lalpha_locations} as the systematic uncertainty on the line peak energies or widths.

\begin{figure}
\begin{center}
\parbox{\linewidth}{
	\includegraphics[width=\linewidth]{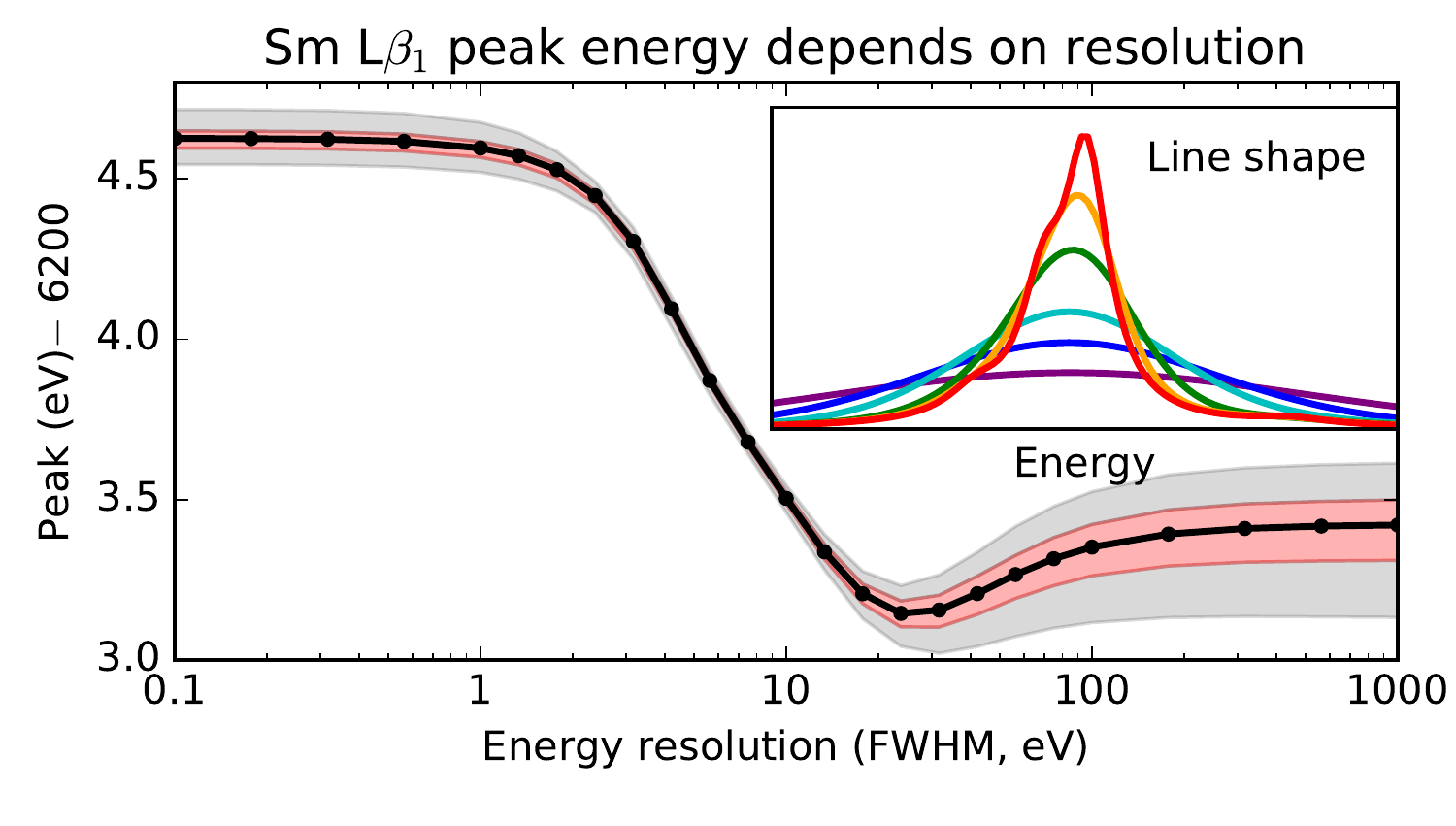}
}
\end{center}
\caption{The energy of the peak as observed by a hypothetical detector with a Gaussian energy-response function will depend on the energy resolution.  The peak of the best-fit Sm L$\beta_1$ line model exceeds 6204.5 eV when the resolution is less then 2\,eV  (FWHM) but is a full 1\,eV lower when the energy resolution is at least 10\,eV.  The black line with dots shows the model that best fits our data; the narrow shaded region shows the 25th to 75th percentile range of 500 simulations in which a 4-component Lorentzian model is fit to similar simulated data, and the wider shaded region shows the 5th to 95th percentile.
The inset shows the line shape over a 60\,eV energy range for resolutions of 0.1, 5, 10, 20, 30, and 50\,eV (top to bottom). }
\label{fig:peak_smear}
\end{figure}

\begin{figure}
\begin{center}
\parbox{\linewidth}{
	\includegraphics[width=\linewidth]{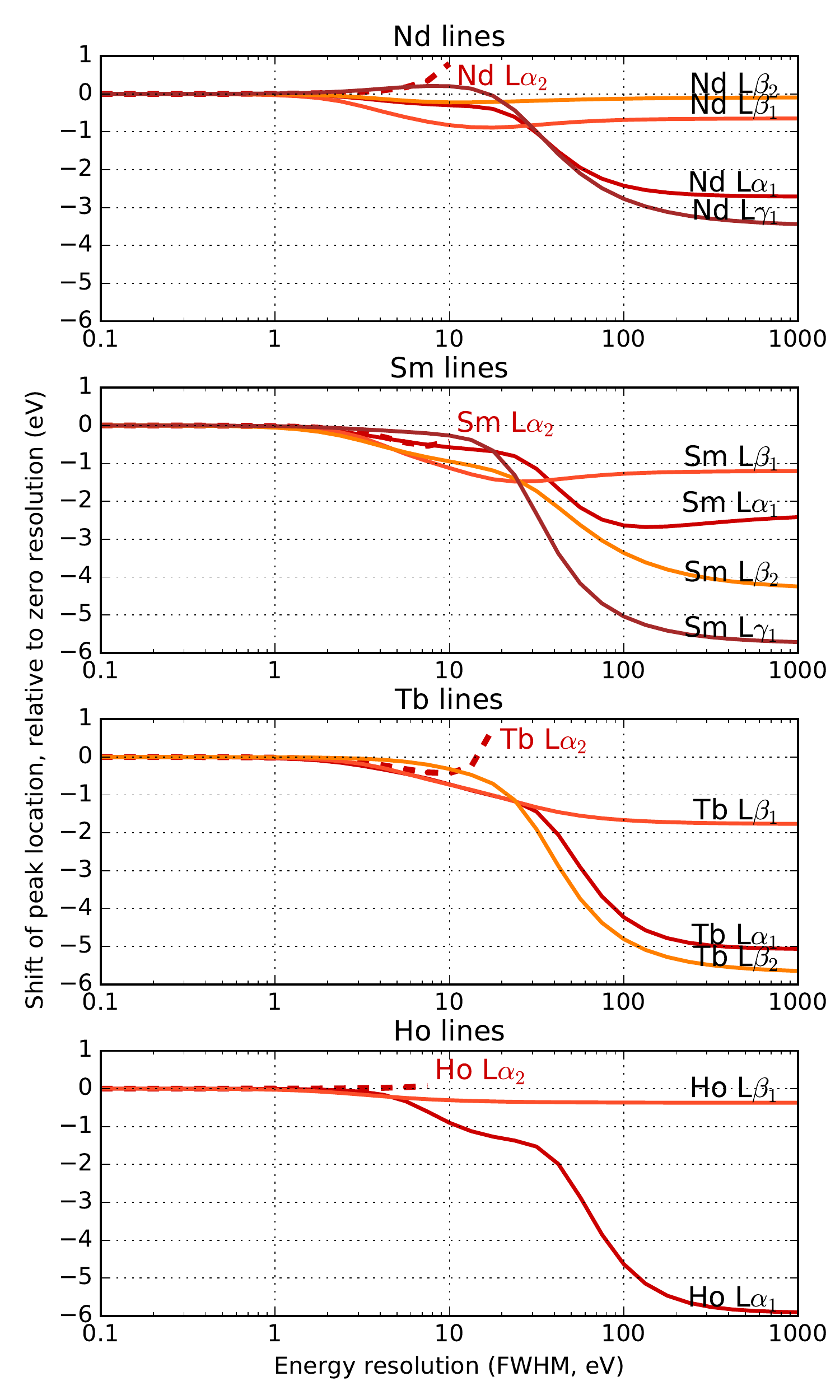}
}
\end{center}
\caption{The shift of the peak in each fluorescence line, as a function of the resolution (FWHM) of a hypothetical instrument with Gaussian energy response. The Ll and L$\beta_3$ lines are symmetric and show no shift, to the level we can measure, so they are omitted. The L$\alpha_2$ lines are shown in dashed lines and only for energy resolutions below approximately 10\,eV; above this level, they cannot be resolved from the L$\alpha_1$ lines. }
\label{fig:peak_smear_all_lines}
\end{figure}

One important effect that can lead to systematic energy offsets is the fact that the value of an asymmetric line's peak energy depends on the energy resolution of the instrument that measures it. We have attempted here to place peaks at the energy where they would be seen in the limit of an ideal instrument with energy resolution much finer than the scale of any features in the intrinsic line shapes. This is essentially a deconvolution problem, and many of the systematic errors in Table~\ref{tab:systematics} follow from the challenges in deconvolving our uncertain instrument response function from the data. For users of a fundamental-parameters database, however, an instrument with limited resolution will produce offsets in the peak energies relative to the peaks presented here. Figure~\ref{fig:peak_smear} shows the Sm L$\beta_1$ peak as an example. An instrument with energy resolution of $\delta E>100$\,eV will find the peak to be a full 1.3\,eV lower than will an instrument with $\delta E<1$\,eV\@.  Figure~\ref{fig:peak_smear_all_lines} shows the estimated size of the shift in the peak for all the lines measured in this work, over a wide range of possible energy resolutions. Most line shapes have pronounced tails or secondary peaks on the low-energy side of the main peak, so that any instrumental broadening tends to reduce the observed energies of the peaks.

\subsection{Discussion of results}

We have measured the fluorescence line energies of 22 lanthanide L lines in the energy range of 4.6\,keV to 7.5\,keV\@. Unlike measurements of an earlier era, these measurements are amenable to thorough analysis of the systematic uncertainties on the absolute-energy scale. We find these uncertainties to be between  0.25\,eV to 0.4\,eV, generally dominated by the uncertainty in the calibration procedure for absolute energies and in the modeling of the intrinsic line shape.

\begin{figure*}
\begin{center}
\parbox{\linewidth}{
	\includegraphics[width=\linewidth]{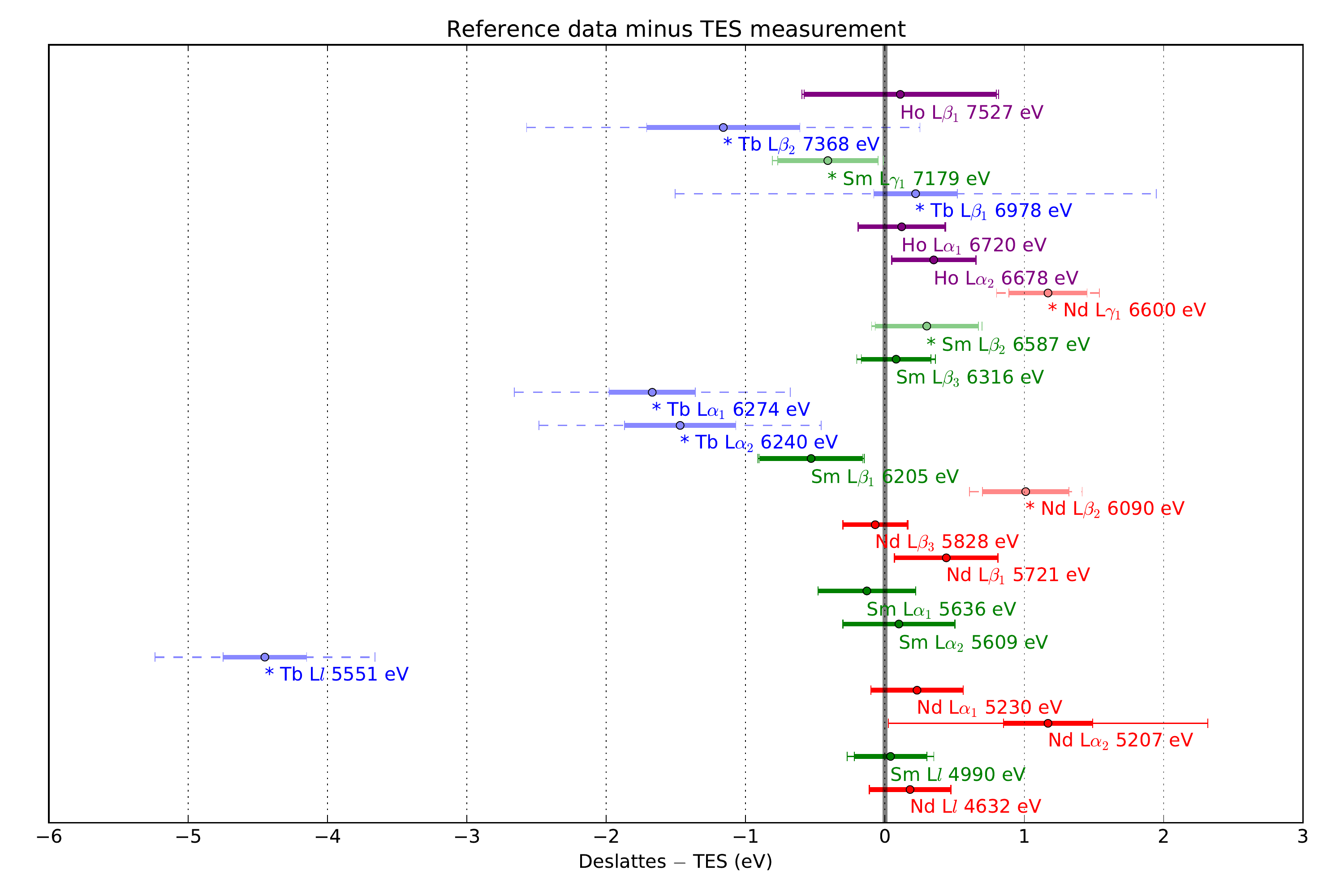}
}
\end{center}
\caption{The energy of the peak given in the reference data minus our results (Tables~\ref{tab:Lbeta_locations} and \ref{tab:Lalpha_locations}). Circles indicate the best estimate of the difference; the thicker, shorter error bars indicate our uncertainties (systematic plus statistical); the longer, thin error bars include both these and the reference data uncertainties, added in quadrature. All reference values that cite T.~Mooney (private communication, 1995) are shown as solid lines; these unpublished data were measured at NIST by a vacuum double-crystal spectrometer. All reference values with only pre-1980 literature citations in \citet{Deslattes:2003} are set off by the use of an asterisk by the element symbol, dashed error bars, and fainter colors. We find complete agreement between the Mooney measurements and our TES-based ones; all discrepancies found here come from the older data.}
\label{fig:reference_compare}
\end{figure*}

Of the sixteen peak energies we have measured for the elements Nd, Sm, and Ho, whose line energies appear in the reference work with uncertainties between 0.03\,eV and 0.26\,eV, we find consistency between our measurement and \citet{Deslattes:2003} for fourteen. The two exceptions are Nd  L$\beta_2$, and L$\gamma_1$, which were not among the neodymium lines measured directly by Deslattes' team; instead, their reference values date to 1977 and earlier~\citep{Bearden:1967tg, Gokhale:1970ve, Shrivastava:1977jd}. The energy differences (reference minus this work) are +1.0\,eV and +1.2\,eV\@. We believe these discrepancies merit follow-up measurements with wavelength-dispersive spectrometers, but our thorough uncertainty analysis gives us confidence in all our results, including these discrepant findings. Figure~\ref{fig:reference_compare} shows the difference between the reference and the TES-based values. The pre-1980 reference data are indicated by fainter colors and asterisks in the figure.

\added{Complete theoretical computations of the line energies for most lines (lines that involve only M-to-L transitions) are provided in~\citet{Indelicato:1998}. These calculations cover all lines measured in this work other than the L$\beta_2$ and L$\gamma_1$ lines (for these lines, references listed in~\citet{Deslattes:2003} give theoretical values). Our measurements of line energies agree with these theoretical results to within 1.5\,eV in all seventeen cases of M-L transitions and in three of five cases of N-L transitions. One of the exceptions, Sm L$\gamma_1$, is discussed below.}

We improve dramatically on the prior work on the energies of five terbium lines, where the primary reference data all appeared in the 1960s or earlier~\citep{Bearden:1967tg}, and of the one modern measurement with an uncharacteristically large uncertainty given by Deslattes as 1.1\,eV: the Nd L$\alpha_2$ line. Although the published uncertainties on terbium line energies were already large, typically 1\,eV or more, it seems likely that the uncertainties were nevertheless underestimated, given the difference of $-4.5$\,eV between Deslattes and this work for the Tb \Ll\ line and $-1.5$\,eV for the \La lines.  Our updated value for the Tb \Ll\ line energy, in particular, resolves a large discrepancy between theoretical calculations and measurement. Deslattes reports $5546.81\pm0.73$\,eV for the measurement (citing~\citet{Bearden:1967tg}) and $5552.7\pm1.5$\,eV for theory~\added{\citep{Indelicato:1998}}. Our \removed{work} \added{measurement} favors the higher, theoretical value. This theory-experiment difference is twice as large as that for any other line measured in this work, save one. For the other large discrepancy, in Sm L$\gamma_1$, our measurement is consistent with the earlier measurement and not with the theory value \added{of 7185$\pm$3\,eV}\@. In summary, we believe we have improved on the energy determination of eight lines out of the 22 lines studied, or nearly 40\,\%.

If we consider only the thirteen lines for which more recent reference data exist, we find that their agreement with the new, TES-based measurements is better than expected. When we combine our uncertainties in quadrature with those of the reference data, the $\chi^2$ statistic of these thirteen differences in peak energy is only 7.1, for which the probability-to-exceed is 0.10.  This low $\chi^2$ could be merely chance, or it could suggest that the uncertainties (here or in the reference data) have been overestimated somewhat (by a factor of approximately 1.5).

The microcalorimeter spectrometer is shown to be capable of measuring intrinsic line widths and line shapes as well as line energies, provided that the intrinsic features are not much smaller than 5\,eV in scale. The analysis of systematics establishes that the typical line's full width at half-maximum is estimated to 0.5\,eV accuracy. The limitations on this measurement are primarily the uncertainties in the TES energy-response function (especially the energy resolution), though the complex and uncertain line shapes can also be a limitation. The line widths are generally close to the theoretical values given by~\citet{Zschornack:2007wu} \added{and by~\citet{Campbell:2001}} particularly for the wider lines ($\Gamma>8$\,eV)\@. The theoretical values for the narrower emission lines (the L$\alpha$, L$\beta_1$, and L$\beta_2$ lines) are 4\,eV to 6\,eV, while the widths we measure are in the range 6\,eV to 8\,eV. \removed{The} \added{Some} theory values are published without uncertainties \removed{and without clarification as to whether satellite line effects are included}, \added{and they omit satellite line effects}; both omissions frustrate detailed comparison.

\subsection{Prospects for future measurements}
\label{sec:prospects}

Future spectrometers are planned to have superior energy linearity and energy resolution, compared to the microcalorimeters used in the current work. Several key improvements have already been established in the laboratory. For one, we have made sensors with larger heat capacity, which leads to a more nearly linear energy response. As expected, initial tests show these sensors improve the performance of smooth approximations to calibration anchor points as described in Section~\ref{sec:calibration}. Specifically, we have demonstrated the reduction of the uncertainties due to the calibration curves (item \ref{syserr:calibration} in the above list) by a factor of $2.0\pm0.4$. 

Better energy-response functions, on the other hand, will improve our estimates of intrinsic line widths and our ability to measure the intrinsic line shape. Energy resolutions of 2.55\,eV (FWHM) at 6\,keV in a fully multiplexed configuration have recently been demonstrated~\citep{Doriese:2015il}. A separate improvement planned for future metrology work is the replacement of bismuth as an absorbing material. A gold absorber has been shown to have a nearly Gaussian energy-response function, thereby eliminating the low-energy tails at only a modest cost in the efficiency of x-ray photon absorption. Simulations show that the elimination of the low-energy tail from the energy-response function and improvement of the Gaussian resolution to 2.6\,eV should reduce by a factor of 2 the uncertainties in both peak energies and widths due to the uncertain resolution (item \ref{syserr:resolution} in Section~\ref{sec:systematics}) and to the uncertain model shape (item \ref{syserr:lineshape}). We are also exploring the possibility that we can expose detectors to tunable, nearly monochromatic radiation sources. Such experiments would allow direct measurement of the TESs' energy-response functions and eliminate one source of uncertainty.

The goals of improved linearity and resolution are often in tension. If this tension cannot be resolved, one approach we will consider for future metrology measurements is to build an array that mixes two TES designs: half the sensors can have better resolution for determination of line shapes, while the other half can have superior linearity for optimal energy calibration results. 
The interpolation of calibration curves and measurement of line shapes have been the most important limitations on the current results, so we anticipate that the sensor improvements described here will enable future measurements with major reductions in systematic uncertainties.

Future arrays are also planned with one thousand or more TESs. Larger arrays will permit measurements of fluorescence lines of lower intensity and will offer even more independent estimates of line energies. Although the array described in this work consists of sensors optimized for the 4\,keV to 8\,keV energy range, the basic TES technology is compatible with measurements over a much broader energy range. We have used TES spectrometers at synchrotrons to measure photon energies as low as 270\,eV~\citep{Ullom:2014er,Fowler:2015MPF}; another spectrometer in which bulk absorbers of tin were attached to the TESs achieved excellent resolution for photons up to 210\,keV~\citep{Bennett:2012kf,Hoover:2013,Winkler:2015}. The technology described in this work can potentially be applied to the estimation of a substantial portion of all x-ray fluorescence line energies and shapes.

\section{Conclusions}
\label{sec:conclusions}

We have used microcalorimeter spectrometers for the first time to deduce the absolute energy of 22 x-ray fluorescence L lines. These measurements, in the energy range of 4.6\,keV to 7.5\,keV, have typical systematic uncertainties of 0.3\,eV and statistical errors that are an order of magnitude smaller. Because these systematic uncertainties are generally dominated by the calibration procedure and the line shape determination, we plan to make future, better measurements with detectors having improved linearity, resolution, and Gaussian response. We find that most line energies agree well with existing reference data. The reference values of some of the measured line energies, particularly lines of terbium, have large uncertainties (1\,eV to 2\,eV), and we reduce these uncertainties to between 0.3\,eV and 0.5\,eV\@. Overall, we believe we have improved on the determination of line energies for eight of the 22 lines measured.

The energy-dispersive measurement technique enabled by microcalorimeters also makes it possible to estimate line widths, provided they are not substantially narrower than the sensors' intrinsic  resolution (here, $\sim$5\,eV). Future spectrometers with improved energy resolution will have corresponding improvements in their ability to measure line widths and shapes. Line widths have not previously been included in some reference data, including the NIST SRD-128 database.

We also note that we have achieved an absolute calibration across a broad energy range with sub-eV accuracy, an accomplishment not previously demonstrated with TESs. In so doing, we have shown that the requirements of future x-ray astronomy missions, such as \emph{Athena}~\citep{Takahashi:2012dh} can be met by microcalorimeters, provided the sensors are comparably linear and the calibration reference energies are suitably dense.

The total time to acquire the data presented here was less than 40\,hours, or less than two hours per fluorescence line. Either the time required or the minimum accessible line intensity will soon be reduced by more than an order of magnitude, as array sizes and sensor speeds improve in the near future.
The current reference data~\citep{Deslattes:2003} contain roughly two hundred K and L lines from 1\,keV to 25\,keV with stated uncertainties exceeding 0.3\,eV, while the more than fifty strong M lines from stable elements do not appear at all in the NIST reference data. With the relatively fast measurement speed enabled by TES microcalorimeters, we plan to apply this technique on many additional fluorescence lines. We also intend to investigate line shifts due to chemical states.

We believe these results merit follow-up measurements with wavelength-dispersive instruments. Such additional measurements would test the uncertainty analysis and enlarge the set of well-measured calibration anchor points in energy ranges where such points are now scarce.

In summary, we have successfully demonstrated a new approach to metrological measurements of x-ray fundamental parameters. Improved measurements of these parameters will be useful across a broad range of work that employs x-ray analytical data. This measurement highlights the capability of superconducting microcalorimeters for metrology.  In the future, we anticipate both improving our measurements within the energy range presented here and expanding the work to cover energies below 4\,keV and above 7.5\,keV\@. 




\begin{acknowledgments}
We acknowledge extensive collaboration with the HEATES exotic atom spectroscopy team, particularly Hideyuki Tatsuno, Shinji Okada, and Shinya Yamada, on matters of broad-band calibration. We also thank Andrew Hoover, Mark Croce, and Michael Rabin at Los Alamos National Laboratory, who have worked with us on calibration strategies in the gamma-ray regime. We thank Jack Wang and Marcus Mendenhall for reviewing this manuscript carefully and Michael Frey and Zydrunas Gimbutas for helpful discussions.
We gratefully acknowledge financial support from the NIST Innovations in Measurement Science Program, and from NASA through the grants ``Providing, enabling and enhancing technologies for a demonstration model of the Athena X-IFU,'' NASA NNG16PT18I, and ``Demonstrating Enabling Technologies for the High-Resolution Imaging Spectrometer of the Next NASA X-ray Astronomy Mission,'' NASA NNH11ZDA001N-SAT. This work was supported by National Research Council Post-Doctoral Fellowships to KM and GO.  CS performed this work under the financial assistance award No. 70NANB15H051 from U.S. Department of Commerce, National Institute of Standards and Technology.    

Contribution of NIST; not subject to copyright.
\end{acknowledgments}

\bibliography{lanthanide_metrology}

\end{document}